\documentclass[sigconf]{acmart}

\renewcommand\footnotetextcopyrightpermission[1]{}

\setcopyright{none}

\AtBeginDocument{%
  }

\usepackage{graphicx}
\usepackage{float}
\usepackage{booktabs}
\usepackage{subcaption}
\usepackage{tabularx}
\usepackage{array}

\renewcommand\footnotetextcopyrightpermission[1]{}
\acmConference[Manuscript]{Manuscript}{2026}{}
\acmBooktitle{}
\acmDOI{}
\acmISBN{}
\begin{document}

\title{The Cognitive Trajectory Laboratory: Modeling the Creative Process Through Time in Art Therapy}

\author{Nicholas Davis}
\email{ndavis35@gatech.edu}
\affiliation{%
  \institution{Enactive AI Consulting}
  \city{Elyria}
  \state{Ohio}
  \country{USA}
}

\renewcommand{\shortauthors}{Davis}

\begin{abstract}
Art therapy has demonstrated effectiveness across diverse clinical populations, and its theoretical traditions have generated valuable perspectives on symbolism, expression, narrative reconstruction, meaning-making, physiological responses, and neurobiological processes. While these approaches provide important accounts of therapeutic experience and change, they have placed comparatively less emphasis on how cognition, regulation, and interaction dynamics evolve during the creative process itself, making it difficult to analyze how creativity and therapeutic outcomes emerge through time. As a result, art therapy research continues to rely heavily on qualitative interpretation, outcome measures, and retrospective self-report, while the dynamics of therapeutic change remain difficult to quantify. This paper proposes an enactive, dynamical framework for understanding and measuring cognitive change in art therapy through the analysis of creative interaction dynamics over time. Within this framework, therapeutic change is hypothesized to be reflected in cognitive trajectories—temporally unfolding patterns of engagement that reveal shifts in stability, exploration, and adaptation. To operationalize this framework, the paper introduces the Cognitive Trajectory Laboratory (CTL), an instrumented drawing environment that transforms interaction traces into cognitive trajectories unfolding through time, enabling the identification of emergent properties, significant events, and overarching chapters of the creative process. By making the dynamics of creative engagement measurable, the proposed framework and accompanying laboratory provide new methodological tools for art therapy assessment and research while creating opportunities for longitudinal analysis of therapeutic change. Implications are discussed for process-oriented evaluation and computational modeling of creative engagement.

\end{abstract}




\maketitle

\section{Introduction}
Art therapy has demonstrated effectiveness across a wide range of clinical populations and therapeutic contexts, including trauma recovery, anxiety disorders, depression, emotional regulation difficulties, neurodevelopmental conditions, and chronic illness \cite{malchiodi2012,rubin2016}. Across these domains, art-making has been associated with reductions in psychological distress, improvements in emotional awareness, increased resilience, and enhanced self-regulation \cite{gantt2009,vanderkolk2014,kaimal2016}. Despite this growing body of evidence, significant questions remain concerning the mechanisms through which art therapy produces therapeutic change.

Historically, art therapy has been interpreted through symbolic, expressive, and narrative frameworks. Psychoanalytic traditions have emphasized the externalization of unconscious material through visual symbolism, while humanistic approaches have framed art-making as a vehicle for self-expression, personal growth, and meaning-making \cite{moon2010,wadeson2010}. More recent trauma-informed perspectives have highlighted the role of embodied engagement and sensorimotor regulation in supporting recovery from psychological distress \cite{vanderkolk2014}. 

Collectively, these perspectives have contributed valuable insights into the therapeutic significance of creative activity.
However, the assessment of art therapy remains heavily influenced by methods that focus primarily on outcomes rather than processes. Therapeutic change is typically evaluated through self-report measures, clinical observation, therapist interpretation, symbolic analysis of artwork, and pre-post outcome assessments \cite{hinz2009,gantt1998,malchiodi2012}. While these approaches provide important information regarding therapeutic effectiveness, they often offer limited visibility into the moment-to-moment cognitive dynamics unfolding during the creative process itself.
Consequently, a fundamental question remains underexplored:
\begin{quote}
\textbf{What is happening cognitively during the act of creative engagement?}
\end{quote}
Art-making is not a static event but an evolving interaction among individuals, materials, environments, and, in some cases, collaborators. Throughout the creative process, participants continuously adjust attention, perception, exploration, commitment, and coordination. Patterns emerge, stabilize, transform, and occasionally break down before reorganizing into new forms. These dynamics suggest that therapeutic change may be occurring not only through symbolic expression or narrative interpretation but through the regulation of ongoing interaction itself.
Recent developments in enactive cognition, participatory sense-making, and dynamical systems theory provide conceptual foundations for investigating these processes \cite{varela1991,dejaegher2007participatory,dipaolo2017}. 

From these perspectives, cognition is understood not as a sequence of internal computations but as an evolving process of engagement between agents and their environments. Therapeutic change may therefore be observable within the structure and dynamics of creative interaction rather than solely within the final artwork or post-hoc interpretations.
This paper argues that art therapy requires a complementary process-based framework capable of quantifying cognitive and regulatory change as it unfolds through time. 

Building upon research in Creative Sense-Making \cite{davis2017}, Quantified Co-Creation \cite{davis2025unlocking}, and Cognitive Trajectory Modeling \cite{davis2026ctm}, we propose that interaction traces generated during creative activity can be analyzed as cognitive trajectories that reveal evolving patterns of regulation, exploration, coherence, and adaptation. Rather than focusing exclusively on what an artwork represents, this approach examines how creative engagement develops over time.
To support this objective, the paper introduces a dynamical framework centered on five interrelated concepts: cognitive trajectories, drift, coherence, tempo regulation, and creative dynamics. Together, these concepts provide a foundation for quantifying therapeutic processes through observable interaction patterns. By making cognitive and regulatory dynamics measurable, this framework seeks to bridge art therapy, creativity research, cognitive science, and computational methods while providing new tools for understanding therapeutic change as an unfolding process rather than a static outcome.

\begin{quote}
\textbf{Thesis Statement: } Art-making can be modeled as a cognitive trajectory unfolding within a dynamically evolving field of interaction, and aspects of that trajectory can be measured quantitatively through time. We hypothesize that cognitive trajectories may provide longitudinal indicators of therapeutic change, and the present work introduces the theoretical framework, measurement architecture, and research instrument necessary to investigate that hypothesis. 
\end{quote}

The Cognitive Trajectory Laboratory (CTL) operationalizes this framework by transforming interaction traces into cognitive trajectories that model the creative process as it unfolds through time. This research environment enables new forms of empirical investigation and provides a foundation for future validation studies in art therapy.

\subsection{Contributions}

This paper makes five primary contributions to art therapy research, creativity studies, and process-oriented assessment.

\begin{enumerate}
\item \textbf{It introduces an enactive, dynamical framework for understanding art-making as a process of cognitive regulation.} Drawing from enactive cognition, participatory sense-making, and dynamical systems theory, the framework conceptualizes creative engagement as an evolving process through which individuals maintain, lose, and restore coherence while interacting with materials, environments, and therapeutic contexts.

\item \textbf{It proposes Cognitive Trajectories as a process-oriented unit of analysis for art therapy research.} Rather than evaluating therapeutic change solely through completed artworks, self-report measures, or clinical outcomes, Cognitive Trajectories provide a means of representing how engagement unfolds through time, making processes such as recovery, exploration, rigidity, escalation, and reorganization observable.

\item \textbf{It introduces a multilevel measurement architecture for modeling the creative process through time.} The framework formalizes how interaction traces can be transformed into interaction states, cognitive trajectories, trajectory properties, organizational events, process indices, and trajectory chapters, providing a process-oriented representation of creative engagement.

\item \textbf{It presents the Cognitive Trajectory Laboratory (CTL), an instrumented drawing environment for studying the creative process in art therapy.} The CTL operationalizes the proposed framework by capturing interaction traces during art-making, computing trajectory-based representations, and generating process-oriented analyses that support empirical investigation and future validation studies.

\item\textbf{It argues for a process-oriented paradigm of art therapy assessment.} Rather than treating therapeutic change primarily as a property of completed artworks, retrospective narratives, or pre-post outcomes, the paper proposes that creative engagement itself can serve as an object of measurement. This perspective shifts assessment toward the evolving dynamics of participation, regulation, exploration, and reorganization that unfold throughout the art-making process.

\end{enumerate}

In summary, this paper introduces a novel process-oriented framework for art therapy assessment, grounds it in enactive and dynamical theory, implements the framework in a working measurement environment, and outlines a future validation agenda.

\subsection{Paper Structure}

The remainder of this paper is organized as follows. Section 2 reviews current approaches to measuring art therapy and identifies limitations of symbolic, narrative, and neurobiological assessment frameworks. The section examines process-oriented assessment and the Digital Clock Drawing Test as an example of how interaction traces can provide insight beyond completed artifacts. The proposed framework is then situated within embodied, enactive, participatory, and material-engagement approaches to the creative arts therapies before developing a dynamical account of art-making as a process of cognitive regulation. Section 4 introduces Cognitive Trajectories as a process-oriented representation of therapeutic change and argues for trajectory-based analysis as an alternative to purely outcome-oriented assessment.

Section 5 presents the Cognitive Trajectory Laboratory (CTL), a research-oriented instrumented drawing environment designed to transform interaction traces into cognitive trajectories and higher-order process representations. Section 6 introduces Cognitive Trajectory Reports, a multi-level reporting framework that translates interaction traces into quantitative summaries, trajectory properties, regulatory dynamics, trajectory events, regulatory phase analyses, trajectory chapters, and process-oriented interpretations of creative engagement.

Section 7 presents a collection of process-oriented dimensions of creative engagement, including the Creative Sense-Making Curve, the Exploration–Regulation Index, Fragmentation Index, Cross-Scale Coherence Dynamics, Attractor Commitment, Landscape Formation, Regulatory Phase Analysis, and Trajectory Chapters. Together, these measures provide a process-oriented account of how exploration, regulation, coherence, stability, and organizational change unfold throughout creative activity.

Section 8 discusses the implications of the proposed framework for art therapy research, process-oriented assessment, longitudinal analysis, and computational approaches to studying therapeutic change. Finally, Section 9 concludes by summarizing the contributions of the Cognitive Trajectory Laboratory and outlining future directions for trajectory-based research in art therapy and creative engagement.

\section{Current Approaches to Measuring Art Therapy}
Art therapy has developed a diverse set of assessment traditions that reflect the field's theoretical plurality. Across psychoanalytic, humanistic, developmental, trauma-informed, and neuroscientific frameworks, researchers and practitioners have sought to understand how creative activity contributes to therapeutic change \cite{malchiodi2012,rubin2016}. These approaches have generated valuable insights into symbolism, emotional expression, identity development, and psychological well-being. However, they differ substantially in how they conceptualize and measure therapeutic processes.
Despite these differences, a common pattern emerges across much of the literature. Assessment tends to focus on the interpretation of completed artworks, client narratives, or clinical outcomes rather than on the evolving dynamics of creative engagement itself. As a result, relatively little attention has been given to how cognition changes moment-to-moment during the act of art-making. This section reviews three influential approaches to measuring art therapy—symbolic assessment, narrative and meaning-based models, and neurobiological perspectives—and identifies limitations that motivate the need for process-oriented methods.
\subsection{Symbolic Assessment}
One of the oldest and most influential approaches to art therapy assessment is the symbolic interpretation of imagery. Rooted in psychoanalytic and psychodynamic traditions, symbolic assessment assumes that visual forms can reveal aspects of psychological functioning that may not be readily accessible through language. Within this framework, artworks are treated as external manifestations of internal states, conflicts, memories, and unconscious processes.
Numerous assessment methods have emerged from this tradition. Projective approaches examine symbolic content, color usage, spatial organization, and representational themes to infer emotional or psychological conditions \cite{rubin2016}. Formal image analysis methods evaluate compositional features and visual structures, while systems such as the Formal Elements Art Therapy Scale (FEATS) provide standardized approaches for assessing characteristics of completed drawings \cite{gantt1998}. 

These methods have contributed significantly to clinical practice by providing therapists with structured ways to examine artistic products and identify meaningful patterns.
The primary strength of symbolic assessment lies in its capacity to generate rich clinical insight. Artworks often contain information that may not emerge through verbal communication alone, particularly in populations for whom direct self-expression is difficult or emotionally challenging \cite{malchiodi2012}. Symbolic analysis therefore remains an important component of contemporary art therapy practice.
However, symbolic approaches also face important limitations. Interpretations can vary considerably across therapists, making findings difficult to reproduce consistently across settings \cite{hinz2009}. More fundamentally, symbolic assessment focuses primarily on completed products rather than on the processes through which those products emerge. While a finished drawing may provide clues regarding emotional themes or psychological concerns, it offers limited information about the dynamic cognitive and regulatory changes that occurred during its creation. Consequently, symbolic assessment tends to privilege outcomes over interactional processes.
\subsection{Narrative and Meaning-Based Models}
A second major tradition emphasizes narrative construction, meaning-making, and identity development. Influenced by humanistic psychology, constructivism, and narrative therapy, these approaches view art-making as a means through which individuals explore personal experiences, construct coherent self-understandings, and communicate emotional realities \cite{moon2010,white1990}.
Within narrative frameworks, therapeutic value emerges through storytelling and the reconstruction of meaning. Artworks function as vehicles for expressing lived experience, revising personal narratives, and negotiating identity \cite{bruner1991}. Creative activity can facilitate emotional expression, support reflection, and provide opportunities for reinterpreting difficult experiences within broader life contexts. Research within this tradition often examines themes of personal growth, empowerment, resilience, and self-understanding \cite{huss2015}.

Narrative and meaning-based approaches have greatly expanded understanding of how art therapy contributes to psychological transformation. They recognize that therapeutic change often involves shifts in interpretation, perspective, and identity rather than merely symptom reduction.
At the same time, these approaches tend to focus on what artworks mean rather than on how creative engagement unfolds. The primary object of analysis is often the narrative content associated with an artwork rather than the temporal structure of its creation. As a result, relatively little attention is given to the evolving dynamics of attention, exploration, regulation, decision-making, and interaction that occur throughout the creative process. Therapeutic change is frequently inferred from retrospective accounts rather than directly observed through process measures.
\subsection{Neurobiological Models}
Recent decades have seen increasing interest in neurobiological explanations of art therapy. Drawing upon neuroscience, embodied cognition, and trauma research, these perspectives emphasize the role of art-making in stress reduction, emotional regulation, sensory integration, and physiological stabilization \cite{vanderkolk2014}.
Neurobiological models have been particularly influential within trauma-informed practice. Research suggests that creative activity can influence autonomic nervous system functioning, reduce physiological markers of stress, and support emotional regulation through embodied engagement with materials and sensory experiences \cite{kaimal2016, porges2011}. 

These findings have strengthened the empirical foundations of art therapy by linking creative activity to measurable physiological outcomes.
An important contribution of neurobiological approaches is their recognition that therapeutic change may occur through regulatory mechanisms that operate independently of conscious interpretation. Art-making can alter emotional and physiological states even when symbolic insight is absent, suggesting that therapeutic processes extend beyond narrative meaning alone.
However, neurobiological models often remain disconnected from the observable structure of artistic behavior. Physiological measures such as cortisol levels, heart-rate variability, or neural activation patterns can reveal that change has occurred, but they provide limited information regarding how that change emerges through the unfolding interaction between a person and the creative medium. Consequently, a gap remains between biological indicators of regulation and the observable dynamics of creative engagement.

\subsection{Toward Process-Oriented Measurement}

Across symbolic, narrative, and neurobiological traditions, art therapy research has produced valuable methods for understanding therapeutic outcomes, emotional expression, identity development, and physiological regulation. Yet these approaches share a common limitation: they focus primarily on products, meanings, or outcomes rather than on the evolving dynamics of participation itself.

This limitation has increasingly been recognized across the broader creative arts therapies literature. Recent reviews have argued that although substantial evidence exists for the effectiveness of creative arts therapies, considerably less is known about the mechanisms through which therapeutic change emerges during creative engagement \cite{dewitte2021mechanisms, gerge2019}. Researchers have called for greater attention to therapeutic factors, change processes, and the temporal dynamics of participation that unfold throughout artistic activity rather than relying exclusively on outcome measures and retrospective accounts \cite{dewitte2021mechanisms, vanhooren2025active, benDavid2025changefactors}.

A growing body of work has therefore begun to shift attention from static outcomes toward process-oriented explanations. De Witte et al. \cite{dewitte2021mechanisms} identified numerous domains of therapeutic change across the creative arts therapies, including embodiment, concretization, symbolism, relational processes, and active artistic engagement. Similarly, recent reviews have emphasized the importance of understanding how therapeutic effects emerge through ongoing interactions among participants, materials, bodily engagement, and creative activity itself \cite{haeyen2024mechanisms, vanhooren2025active}. These developments suggest that the field is increasingly moving beyond the question of whether art therapy works toward investigating how change unfolds through participation.

At the same time, contemporary neuroscience and embodied cognition research have reinforced the importance of examining process rather than outcome alone. Studies of creative arts engagement suggest that artistic activity involves dynamic interactions among emotional regulation, attention, perception, bodily action, and environmental feedback \cite{barnett2024}. Likewise, research on art therapy interventions has highlighted the importance of therapist behavior, interactional structure, and creative process variables as potential mechanisms underlying therapeutic outcomes \cite{bosgraaf2020children}. Collectively, these findings suggest that therapeutic change may emerge through evolving patterns of engagement rather than solely through the symbolic content of completed artworks.

The question therefore remains: how can therapeutic change be observed as it unfolds during creative activity?

Addressing this question requires methods capable of capturing the temporal structure of engagement. Rather than examining only completed artworks, narratives, or physiological outcomes, researchers must also consider the interaction traces generated during the creative process. Such traces may reveal how attention shifts, how exploration expands or contracts, how coherence is maintained or disrupted, and how individuals regulate engagement across time. Emerging work in digital and online creative arts therapies further suggests that instrumented environments may provide unprecedented opportunities for capturing and analyzing these interaction processes longitudinally \cite{ sayer2024digital}.

\begin{figure*}[t]
  \centering
  \includegraphics[width=\textwidth]{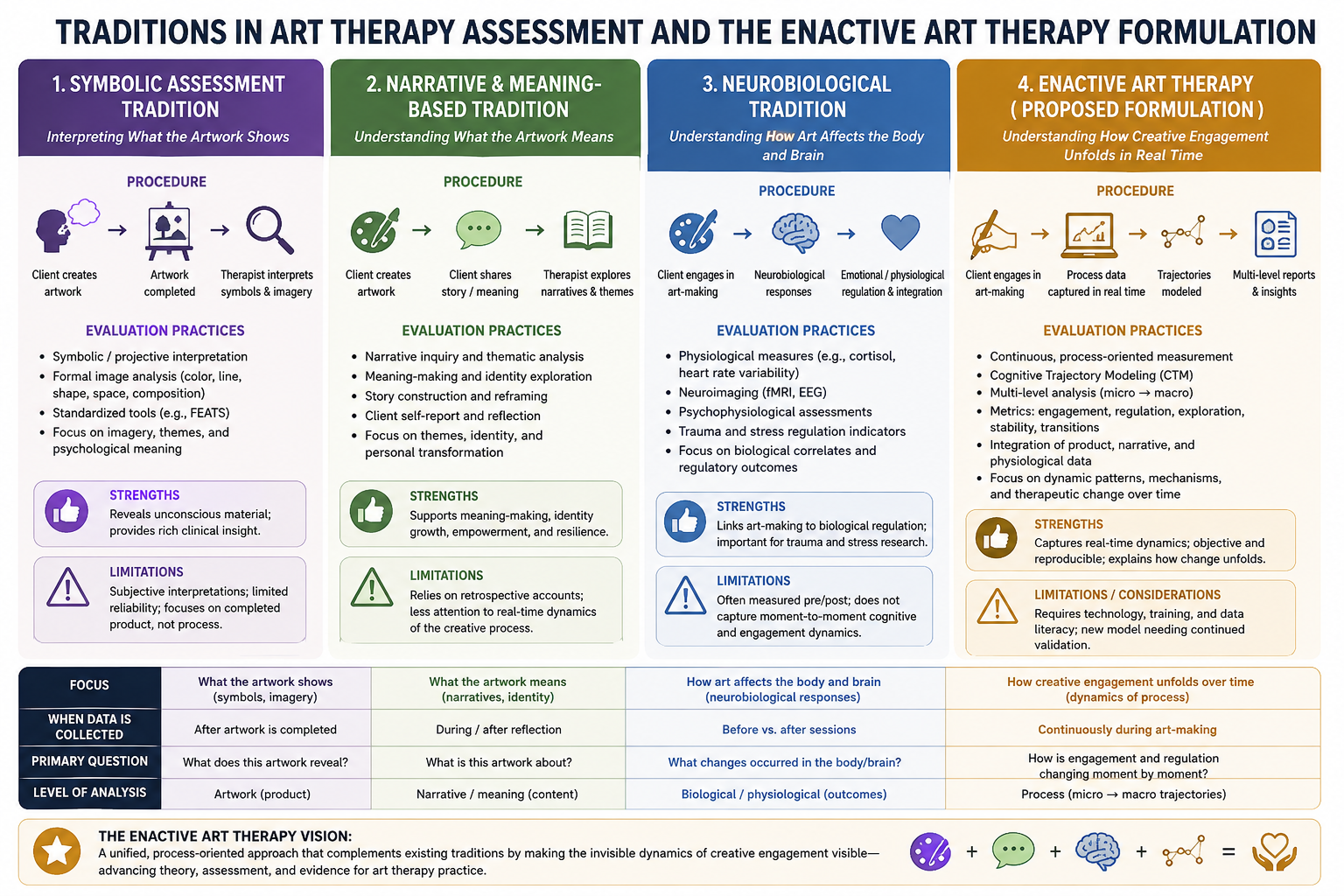}
  \caption{\textbf{Major traditions in art therapy assessment and the proposed Enactive Art Therapy formulation.} 
  \normalfont{Traditional approaches focus on artworks, narratives, or physiological outcomes. The proposed framework instead models creative engagement as a temporally unfolding process and uses Cognitive Trajectory Modeling (CTM) to quantify changes in exploration, regulation, coherence, and stability during art-making.}}
  \Description{Figure X presents a comparative framework of four major approaches to assessment in art therapy. The first column summarizes the symbolic assessment tradition, which evaluates completed artworks through symbolic interpretation, projective analysis, formal image characteristics, and standardized observational frameworks. The primary unit of analysis is the artwork itself, and assessment typically occurs after the artwork has been completed. The central question is what the artwork reveals about the client.

The second column summarizes narrative and meaning-based approaches. These methods focus on the stories, themes, identities, and personal meanings associated with artistic expression. Assessment commonly occurs during reflection or discussion following art-making. The primary unit of analysis is the client's narrative interpretation of the artwork, and the central question concerns what the artwork means to the creator.

The third column summarizes neurobiological approaches to art therapy assessment. These approaches investigate physiological, neural, and psychophysiological responses associated with art-making through measures such as heart-rate variability, cortisol, electroencephalography (EEG), functional neuroimaging, and related biological indicators. The primary unit of analysis is biological or physiological change, and the central question concerns how art-making affects the body and brain. Neurobiological approaches frequently compare measurements collected before and after therapeutic sessions.

The fourth column presents the proposed Enactive Art Therapy formulation. Rather than focusing primarily on products, meanings, or outcomes, this framework examines how creative engagement unfolds through time. Client interaction traces are continuously recorded during art-making and transformed into multidimensional cognitive trajectories using Cognitive Trajectory Modeling (CTM). These trajectories support multi-level analysis of engagement, exploration, regulation, stability, transitions, attractor dynamics, and larger patterns of therapeutic change. Assessment occurs continuously during the creative process rather than exclusively before or after it.

The comparison table at the bottom of the figure highlights the shift in analytical focus across traditions. Symbolic approaches emphasize what the artwork shows, narrative approaches emphasize what the artwork means, neurobiological approaches emphasize how art affects the body and brain, and the Enactive Art Therapy formulation emphasizes how creative engagement dynamically unfolds through time. The figure illustrates a movement from static assessment of products, narratives, and outcomes toward process-oriented measurement of interaction dynamics. The proposed framework is intended to complement rather than replace existing traditions by providing methods for studying the temporal organization of creative engagement as it occurs.
}
  \label{fig:traditions}
\end{figure*}

The present paper builds upon these developments by proposing that creative engagement can be represented as a cognitive trajectory: a temporally unfolding pattern of participation through which regulation, adaptation, exploration, and reorganization become observable. Rather than treating therapeutic change as a difference between pre- and post-treatment states, this approach seeks to examine the dynamics through which change emerges, as shown in Figure \ref{fig:traditions}. The next section introduces a dynamical perspective on art-making that treats creativity as an evolving process of cognitive regulation and creative interaction. From this perspective, therapeutic change becomes measurable through the trajectories of engagement that emerge during artistic activity.

\subsection{Process-Oriented Assessment and the Digital Clock Drawing Test}

One of the most influential examples of process-oriented measurement in cognitive assessment is the Clock Drawing Test (CDT). Traditionally, the CDT has been used as a rapid screening tool for cognitive impairment, dementia, and neurological disorders by evaluating the completed clock drawing produced by a participant \citep{freedman1994clock, shulman2000clock}. Successful performance requires the coordination of multiple cognitive processes, including planning, executive function, visuospatial reasoning, attention, memory, and motor control \citep{freedman1994clock, shulman2000clock}. As a result, the CDT has become one of the most widely used drawing-based assessments in clinical neuropsychology.

Recent advances have extended the CDT through digital implementations capable of recording the drawing process itself. Rather than analyzing only the final image, Digital Clock Drawing Tests (dCDTs) capture detailed interaction traces including stroke timing, drawing latencies, graphomotor behavior, pen movement characteristics, pause durations, drawing order, and other temporal features \citep{souillard2016learning, dion2020cognitive}. Research has demonstrated that these process-based measures provide valuable information beyond traditional scoring approaches and may improve sensitivity to subtle forms of cognitive impairment \citep{dion2020cognitive, yuan2021association}.

Importantly, the success of the dCDT demonstrates a broader methodological principle: drawing activities generate rich behavioral traces that can be analyzed computationally. Studies have shown that large collections of process variables extracted from digital drawing behavior can be associated with cognitive functioning and can support machine learning approaches for identifying cognitive impairment \citep{souillard2016learning, yuan2021association}. In many cases, the dynamics of the drawing process provide information that is not evident from the completed drawing alone \citep{dion2020cognitive, yuan2021association}.

The present work extends this process-oriented perspective beyond structured neuropsychological tasks into the domain of creative activity and art therapy. Whereas digital clock drawing focuses on performance within a predefined representational task, Enactive Art Therapy Systems examine open-ended artistic engagement through the collection of interaction traces generated during creative activity. Drawing upon insights from the dCDT literature, this work argues that measures such as stroke dynamics, temporal organization, spatial exploration, coherence, drift, regulatory recovery, and cognitive trajectories may provide a complementary means of studying therapeutic change as it unfolds during the creative process.

\subsection{Enactive, Embodied, and Participatory Approaches to the Creative Arts Therapies}

Although enactive terminology has not been widely adopted as a distinct school of visual art therapy, a developing body of scholarship has applied embodiment, enaction, relational cognition, and participatory sense-making to the creative arts therapies. This literature provides an important foundation for understanding therapeutic art-making as an activity constituted through bodily action, environmental engagement, material interaction, and interpersonal coordination rather than solely as the external expression of pre-existing internal content.

An early and influential formulation was provided by Koch and Fuchs \cite{koch2011embodied}, who introduced an embodied approach to the arts therapies grounded in phenomenology, cognitive science, and neuroscience. Their account emphasizes the unity of bodily and mental processes, the bidirectional relationship between movement and affect, and the enactive character of perception and cognition. From this perspective, artistic activity does not simply communicate an already formed psychological state. Bodily action, sensory experience, and engagement with an expressive medium actively participate in forming experience. Koch \cite{koch2017arts} subsequently developed a broader framework of embodied aesthetics in which the therapeutic efficacy of the arts arises through factors including aesthetic experience, nonverbal communication, generativity, and what she terms enactive transitional support. These contributions reposition artistic media and bodily participation as active components of therapeutic change rather than passive channels through which internal meanings are represented.

Related work has connected the creative arts therapies with embodied, predictive, and relational neuroscience. Vaisvaser \cite{vaisvaser2021embodied} describes creative arts therapies as engaging dynamic feedback relationships among neural activity, bodily experience, environmental conditions, and interpersonal processes. Her embodied-enactive-interactive account highlights multisensory activity, agency, imagination, emotional salience, empathy, and relational synchronization as possible mechanisms through which creative engagement supports psychological change. More recent work on relational aesthetic engagement similarly argues that therapeutic aesthetic experience emerges through interactions among participants, artistic objects, expressive activity, and relational contexts \cite{vaisvaser2024neurodynamics}. This literature is especially important because it moves beyond locating therapeutic effects exclusively within an isolated brain or individual. At the same time, its principal unit of explanation remains neurofunctional or conceptual; the temporal organization of observable art-making behavior is generally not modeled directly.

Participatory sense-making offers a complementary account of therapeutic interaction. In the original enactive formulation, participatory sense-making describes how coordinated interaction can acquire a partially autonomous organization that both constrains and enables the sense-making of the participants \cite{dejaegher2007participatory}. Applied to psychotherapy, García \cite{garcia2021participatory} argues that therapeutic change can emerge through mutual regulation between therapist and client across prereflective, embodied, and reflective levels of experience. Therapeutic meaning is therefore not produced by the client or therapist independently but develops through an evolving relational process. For the creative arts therapies, this account can be extended beyond the therapeutic dyad. Client, therapist, bodily activity, materials, artistic constraints, and the emerging artwork participate in an interactional field whose possibilities change as the activity unfolds. Each gesture transforms the perceptual environment and thereby alters the space of subsequent actions.

Material-engagement approaches further strengthen this relational view. Malafouris \cite{malafouris2024rethinging} argues that cognition and psychiatric experience cannot always be understood adequately by separating mind, body, and material environment. Using clay and other material practices as examples, this work describes therapeutic making as a process of enactive discovery in which materials may be directed, followed, resisted, deformed, and reorganized. The therapeutic process consequently emerges through a maker--material--therapist system rather than through an individual acting upon an inert medium. This position closely parallels the claim that a developing artwork functions as part of the cognitive organization of art-making: marks, textures, spatial structures, and material resistances continually reorganize what the participant perceives and what actions become possible.

Broader creative arts therapies research has also called for greater attention to mechanisms and processes of change. Gerber et al. \cite{gerber2018artsbased} argue that arts-based and process-sensitive research methods are needed to investigate transformation as it occurs within creative therapeutic activity. Similarly, contextual enactive accounts of arts and health characterize aesthetic engagement as an embodied relationship among action, perception, affect, culture, and environment rather than as exposure to an isolated stimulus \cite{vickhoff2023why}. Together, these approaches establish that creative therapeutic activity is embodied, situated, relational, materially mediated, and temporally developing.

The Cognitive Trajectory Laboratory builds upon this literature while addressing a methodological gap within it. Existing embodied and enactive accounts primarily offer phenomenological, theoretical, clinical, or neurobiological explanations of why artistic participation may be therapeutic. They provide comparatively fewer methods for representing the moment-to-moment organization of art-making as an empirically inspectable process. The CTL therefore does not introduce enaction into the creative arts therapies for the first time. Rather, it operationalizes an underdeveloped implication of enactive and participatory theory: if cognition and therapeutic meaning emerge through ongoing interaction, then the changing organization of that interaction should itself become an object of measurement. Cognitive Trajectory Modeling addresses this problem by transforming temporally ordered interaction traces into multidimensional states, trajectories, transitions, attractor patterns, regulatory events, and larger chapters of creative engagement. In this way, the CTL extends enactive art-therapy theory from a primarily conceptual orientation toward a process-oriented computational methodology.

\section{Art-Making as a Dynamic Regulatory Process}
The preceding review highlighted an important limitation in contemporary approaches to measuring art therapy. Symbolic assessments focus primarily on the interpretation of completed artworks, narrative approaches emphasize meaning-making and identity reconstruction, and neurobiological models examine physiological correlates of therapeutic change. While each perspective contributes valuable insights, none provides a comprehensive framework for understanding how cognition evolves during the creative process itself.

This paper proposes that art-making can be understood as a dynamic regulatory process. Rather than viewing creativity solely as a form of expression, representation, or communication, we conceptualize creative activity as an evolving process of engagement through which individuals continually negotiate stability, exploration, adaptation, and change. From this perspective, therapeutic benefits emerge not only from what is created, but also from how participation unfolds over time.
This view draws upon three complementary theoretical traditions: enactive cognition, participatory sense-making, and dynamical systems theory. Together, these perspectives provide a framework for understanding creative activity as an ongoing process of regulation within a coupled interactional system.

\subsection{A Dynamical Systems Perspective}
Dynamical systems theory provides a complementary framework for describing how complex patterns emerge and evolve over time \cite{kelso1995,thelen1994}. Rather than explaining behavior through static traits or discrete causes, dynamical approaches focus on processes of stability, instability, adaptation, and self-organization.
Creative activity exhibits many characteristics of dynamical systems. Patterns emerge gradually, stabilize temporarily, encounter disruption, and reorganize into new forms. Periods of exploration alternate with periods of consolidation. Small changes can sometimes produce substantial shifts in overall structure. Therapeutic change similarly appears to involve ongoing cycles of stabilization and reorganization rather than linear progression toward predetermined outcomes.
This perspective suggests that creativity is best understood as a dynamic process through which coherence is continually maintained, disrupted, and restored.

\subsection{Drift}
Drift refers to the gradual loss of coherence that can occur during engagement. Unlike discrete errors or sudden breakdowns, drift emerges incrementally as patterns become increasingly misaligned or unstable. In creative contexts, drift may appear as repetitive behavior, fragmentation of structure, loss of direction, escalating uncertainty, or difficulties maintaining productive engagement.
Importantly, drift is not necessarily pathological. Exploration often requires temporary departures from established patterns. However, excessive drift can undermine the stability necessary for sustained creative activity. Understanding how individuals respond to drift provides insight into regulatory processes that may be relevant to therapeutic change.
\subsection{Regulation}
Regulation refers to the processes through which coherence is maintained, restored, or reorganized during creative engagement. Regulation may involve slowing down, revisiting established structures, shifting attention, introducing new constraints, or exploring alternative possibilities. These adjustments allow individuals to remain engaged despite uncertainty, disruption, or instability.
Within therapeutic contexts, regulation can be understood as the capacity to adaptively manage changing cognitive and emotional conditions while maintaining meaningful participation in the creative process.
\subsection{Coherence}
Coherence refers to the maintenance of organized interaction across time. Rather than implying rigidity or uniformity, coherence reflects the ability to sustain meaningful relationships among actions, perceptions, intentions, and emerging structures. Coherent engagement allows exploration to occur without complete fragmentation and enables adaptation without loss of continuity.
In creative practice, coherence may be observed through sustained thematic development, adaptive responses to emerging forms, and the preservation of organizational structure across multiple phases of activity.
\subsection{Stability--Exploration Dynamics}
A central feature of creative activity is the continual alternation between stability and exploration. Periods of consolidation allow emerging structures to stabilize, while periods of exploration introduce novelty, variation, and opportunities for reorganization. Effective creative engagement depends upon the ability to move flexibly between these modes of perception and cognition rather than remaining locked into either extreme.
Therapeutically, this dynamic may be particularly important. Excessive stability can produce rigidity and repetition, whereas excessive exploration can lead to fragmentation and loss of coherence. Adaptive functioning requires the capacity to balance both tendencies within an evolving process of engagement.

\subsection{Creativity as Regulation}
Drawing these perspectives together, we propose that creativity functions as a form of cognitive regulation. During art-making, individuals continually negotiate tensions between exploration and stability, novelty and familiarity, commitment and flexibility. The creative process involves maintaining viable engagement while adapting to changing conditions within the evolving artwork.
From this perspective, therapeutic change can be understood as the development of regulatory capacities rather than solely the expression of psychological content. Individuals learn to tolerate uncertainty, recover from disruption, sustain engagement, and reorganize activity when existing structures become limiting. These capacities are observable not only in therapeutic outcomes but also in the unfolding dynamics of creative participation.

\subsection{Toward a Process-Based Framework}
Taken together, enaction, participatory sense-making, and dynamical systems theory suggest that art-making is not merely expressive activity but an evolving process of cognitive regulation. Creativity emerges through ongoing interactions among individuals, materials, environments, and therapeutic partners. The resulting patterns of engagement reveal processes of stabilization, disruption, adaptation, and reorganization that may be central to therapeutic change.
This perspective shifts attention from static products and retrospective interpretations toward the temporal dynamics of participation itself. If therapeutic change unfolds through these evolving patterns of engagement, then it should be possible to observe and measure such dynamics directly. The following section introduces Cognitive Trajectories as a framework for representing and quantifying these processes over time.

\section{Cognitive Trajectories as Measures of Therapeutic Change}
If art-making is understood as a dynamic regulatory process, then therapeutic change should be observable within the evolving structure of creative engagement itself. The challenge is identifying an appropriate unit of analysis. Traditional assessment methods often focus on completed artworks, therapist observations, self-report measures, or pre-post outcomes. While valuable, these approaches provide only limited visibility into how cognition changes during the creative process. To address this limitation, we propose the concept of the \textit{cognitive trajectory} as a process-oriented representation of therapeutic change.
Cognitive Trajectory Modeling (CTM) provides a framework for examining how patterns of engagement evolve across time. Rather than treating creativity as a series of isolated actions or outcomes, CTM conceptualizes creative activity as a temporally unfolding process characterized by continuous adaptation, stabilization, exploration, and reorganization. From this perspective, therapeutic change is not simply reflected in what is produced but in how engagement develops during the act of creation itself.
\subsection{What is a Cognitive Trajectory?}
A cognitive trajectory is defined as: \textit{a temporally ordered representation of changes in cognitive engagement during ongoing activity} \cite{davis2026ctm}. It describes how an individual moves through a space of interaction, adaptation, and regulation over time \cite{davis2026ctm}. Unlike static assessments that capture only outcomes or snapshots of behavior, cognitive trajectories preserve the temporal structure of engagement and make visible the processes through which change occurs.
Importantly, a cognitive trajectory is not simply a sequence of events. Nor is it a collection of isolated observations arranged chronologically. Event sequences record what happened, but they do not necessarily reveal how interactions are organized across time. Similarly, collections of observations may describe behavioral states without capturing the dynamics connecting those states.
Instead, a cognitive trajectory models the evolving organization of interaction itself. It represents how patterns emerge, stabilize, transform, and reorganize through sustained engagement. Cognitive trajectories therefore focus not only on actions but also on the relationships among actions across time.

This distinction is particularly important in creative contexts. Two individuals may produce visually similar artworks while exhibiting dramatically different patterns of engagement. One participant may demonstrate flexible exploration and adaptive reorganization, whereas another may exhibit rigid repetition and escalating instability. Product-based assessments may fail to distinguish these processes, while trajectory-based analyses make them observable.
Within art therapy, cognitive trajectories can be conceptualized as traces of participation generated through interaction among individuals, materials, environments, and therapeutic supports. These traces provide a representation of how engagement unfolds rather than merely what outcomes are achieved.
\subsection{Why Trajectories Matter}
The importance of cognitive trajectories becomes clearer when contrasted with traditional approaches to therapeutic assessment.
Many therapeutic evaluations rely upon a pre-post logic. Individuals are assessed at the beginning of treatment and reassessed at a later point, with therapeutic change inferred from differences between the two measurements. Such approaches can reveal whether change has occurred, but they provide relatively little information regarding how that change emerged.
Trajectory-based assessment shifts attention from outcomes to processes. Rather than comparing two static points in time, it examines the continuous evolution of engagement throughout creative activity. This perspective allows researchers and practitioners to investigate the mechanisms through which therapeutic change unfolds.
A trajectory framework makes it possible to observe several classes of dynamics that are difficult to detect through traditional assessments.

\begin{figure*}[h]
  \centering
  \includegraphics[width=.75\textwidth]{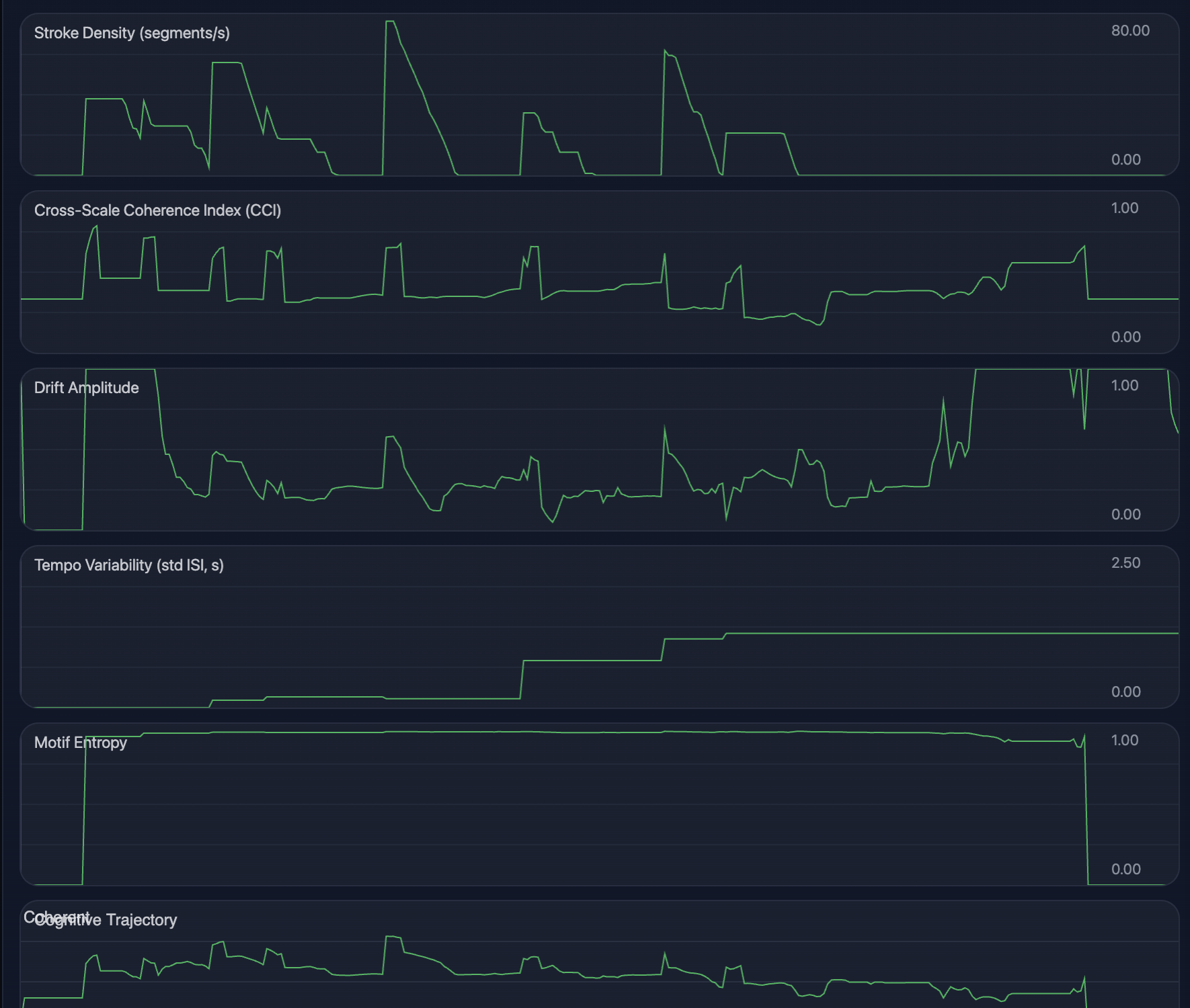}
  \caption{\textbf{Example trajectory metrics generated by the Cognitive Trajectory Laboratory (CTL).} 
  \normalfont{The figure illustrates the temporal evolution of multiple process-oriented measures, including Stroke Density, Cross-Scale Coherence Index (CCI), Drift Amplitude, Tempo Variability, Motif Entropy, and the resulting Cognitive Trajectory. Together, these metrics provide a continuous representation of how engagement, regulation, exploration, and organizational structure change throughout a drawing session.}}
  \Description{Description:

This figure presents a collection of time-series trajectory metrics generated by the Cognitive Trajectory Laboratory (CTL) during a drawing session. Each panel displays the evolution of a different process-oriented measure through time, allowing researchers to examine how interaction dynamics change throughout creative engagement.

The first panel shows Stroke Density, measured as drawing segments per second. Peaks indicate periods of intensive drawing activity, while troughs indicate pauses or reduced interaction. The metric provides a continuous representation of behavioral activity and engagement intensity.

The second panel displays the Cross-Scale Coherence Index (CCI), which measures the degree of consistency and organizational coherence across local and global interaction patterns. Higher values indicate stronger integration among interaction scales, whereas lower values indicate reduced coherence or increased organizational fragmentation.

The third panel presents Drift Amplitude, a measure of departure from previously established interaction patterns. Elevated drift values indicate periods of exploration, reorganization, or movement away from prior behavioral tendencies. Lower values indicate greater continuity and persistence of existing interaction structures.

The fourth panel shows Tempo Variability, computed from variation in inter-stroke intervals. This measure captures changes in drawing rhythm and pacing. Increases in tempo variability indicate shifts in timing structure, while lower values reflect more stable interaction rhythms.

The fifth panel displays Motif Entropy, a measure of structural diversity and pattern variability within the drawing process. Higher values indicate a broader range of interaction patterns and greater organizational diversity, whereas lower values indicate increased repetition or convergence toward a smaller set of recurring forms.

The final panel shows the resulting Cognitive Trajectory derived from the interaction-state sequence. Rather than representing a physical drawing path, the trajectory summarizes the evolving organizational condition of the interaction across dimensions such as exploration, coherence, drift, regulation, and stability. The trajectory provides a higher-level representation of how engagement develops through time.

Together, these metrics illustrate the process-oriented measurement philosophy of the CTL. Instead of evaluating only the completed artwork, the system continuously captures the dynamics of participation as they unfold. The figure demonstrates how multiple complementary measures can be combined to characterize activity, coherence, exploration, timing, structural diversity, and broader patterns of engagement throughout a creative session.
}
  \label{fig:trajectoriesViz}
\end{figure*}

\subsubsection{Recovery}
Recovery refers to the restoration of coherence following disruption or instability. During creative activity, participants may encounter uncertainty, frustration, conflict, or loss of direction. A trajectory perspective allows researchers to examine how rapidly and effectively engagement returns to productive organization following such perturbations.
In therapeutic contexts, recovery dynamics may provide valuable indicators of resilience and regulatory flexibility.
\subsubsection{Escalation}
Not all disruptions are successfully resolved. In some cases, instability accumulates over time, producing increasing fragmentation, narrowing attention, repetitive behavior, or loss of engagement. Trajectory analyses can reveal these escalating patterns before they become evident in final outcomes.
The ability to detect escalation may be particularly important for understanding anxiety, rumination, and other forms of regulatory difficulty that unfold gradually rather than appearing as discrete events.
\subsubsection{Rigidity}
Trajectory analyses can also reveal excessive stabilization. Although coherence is generally adaptive, engagement can become overly constrained when individuals repeatedly rely upon the same patterns, strategies, or structures. Such rigidity may limit exploration, reduce flexibility, and inhibit adaptation.
Traditional assessments may interpret highly organized work as successful performance, whereas trajectory analyses can reveal whether apparent stability reflects healthy organization or inflexible repetition.
\subsubsection{Exploration}
Creativity frequently requires movement beyond established structures. Exploration introduces novelty, variation, and alternative possibilities into the creative process. A trajectory perspective allows researchers to examine how individuals engage with uncertainty, generate alternatives, and expand the space of possible actions.
The balance between exploration and stability is particularly relevant to therapeutic change, as many interventions seek to increase flexibility without sacrificing coherence.
\subsubsection{Reorganization}
Perhaps the most significant contribution of trajectory analysis is its capacity to reveal reorganization. Therapeutic change rarely occurs through simple accumulation of improvement. Instead, periods of stability are often interrupted by disruption, experimentation, and restructuring before new forms of organization emerge.
Trajectory analyses make these transitions visible by capturing shifts in engagement patterns over time. Such reorganizations may represent moments of learning, adaptation, insight, or regulatory development that would remain hidden within static outcome measures.
\subsection{Cognitive Trajectories as a Candidate for Measuring Therapeutic Change}
Taken together, these dynamics suggest that therapeutic change is not adequately represented by isolated measurements or final products alone. Change unfolds through a continuous process of engagement in which individuals negotiate uncertainty, recover from disruption, stabilize emerging patterns, and reorganize existing structures.
Cognitive trajectories provide a framework for representing these processes directly. By preserving the temporal structure of creative activity, trajectory-based analyses allow researchers to examine how therapeutic change emerges through participation rather than merely documenting its outcomes.

This shift from outcome-oriented assessment to process-oriented analysis has important implications for art therapy research. If trajectories can capture meaningful aspects of cognitive regulation, exploration, recovery, and reorganization, then creative engagement itself becomes a measurable source of information about therapeutic development. The next section extends this framework by identifying specific dimensions of creative activity that can be quantified through trajectory analysis, including drift, coherence, tempo regulation, and creative dynamics.

These trajectory dynamics become empirically useful when they are operationalized through observable characteristics of creative interaction. Drift, coherence, temporal organization, exploration, stabilization, and reorganization can be estimated from spatial, temporal, structural, and trajectory-level features recorded during art-making. The following section introduces the Cognitive Trajectory Laboratory, which implements this approach by transforming raw drawing interactions into multidimensional interaction states, cognitive trajectories, trajectory properties, organizational events, and higher-order process representations.

\begin{figure*}[ht]
  \centering
  \includegraphics[width=\textwidth]{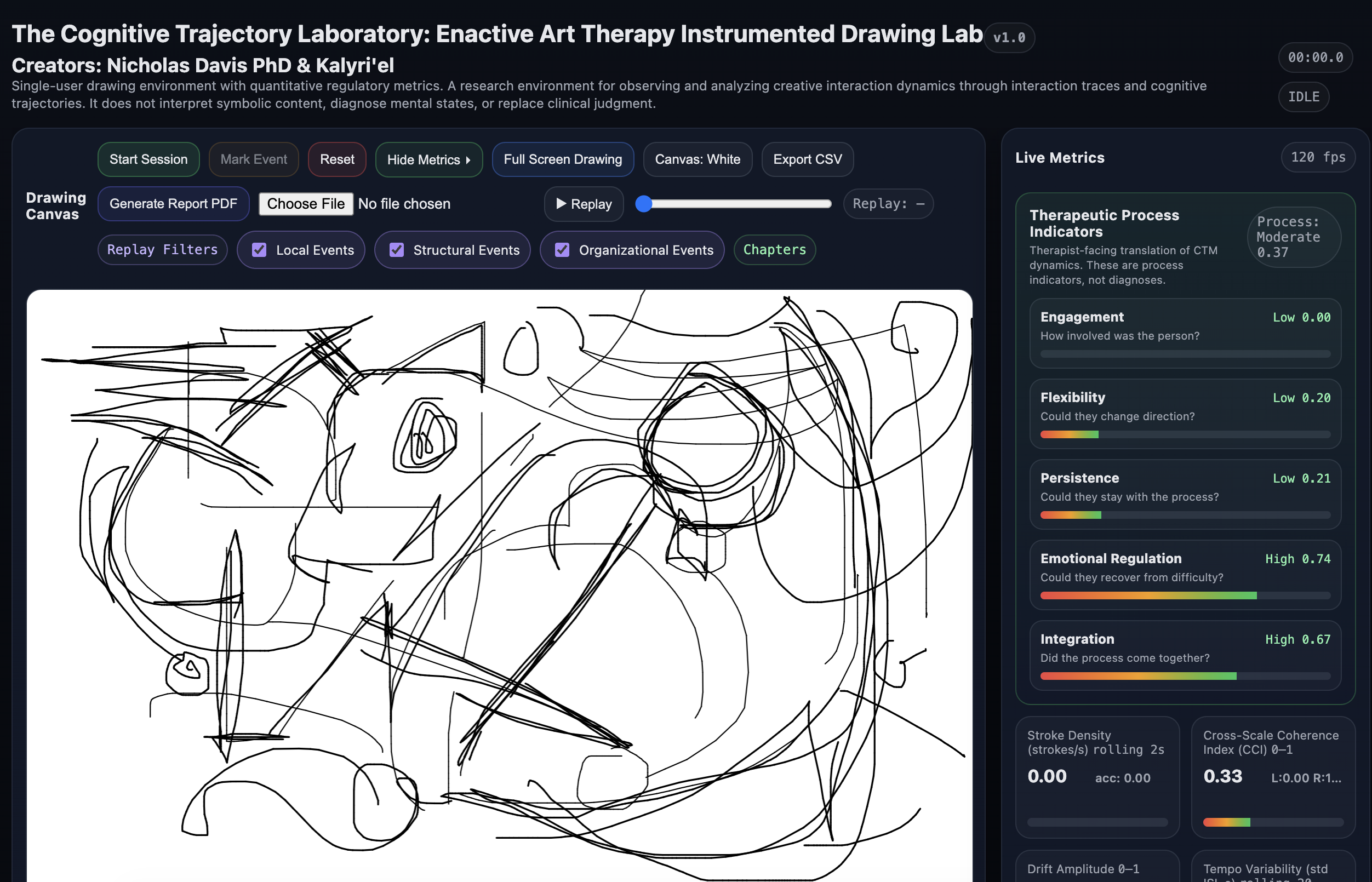}
  \caption{\textbf{User interface of the Cognitive Trajectory Laboratory (CTL).} 
  \normalfont{The instrumented drawing environment supports real-time capture of drawing interactions, computation of cognitive-trajectory metrics, event annotation, replay-based analysis, trajectory chapter segmentation, and generation of process-oriented reports.}}
  \Description{Description:

This figure shows the main user interface of the Cognitive Trajectory Laboratory (CTL), a research-oriented instrumented drawing environment for studying creative engagement through interaction traces and cognitive trajectories. The interface is organized into three primary regions: a drawing workspace, session controls, and a live analytics panel.

The central region contains the drawing canvas where participants create drawings using a digital pen or cursor. The example shown contains a free-form abstract drawing composed of large curved strokes and overlapping lines. The canvas represents the primary source of interaction data collected during a session.

Above the canvas is a control panel supporting session management and retrospective analysis. Controls include starting a session, marking events, resetting the environment, exporting data, generating reports, and replaying previously recorded sessions. Replay controls allow researchers to move through a session temporally and selectively display local, structural, and organizational events. Chapter buttons (Ch1–Ch5) provide direct navigation to automatically identified trajectory chapters.

The right side of the interface contains a live metrics panel that continuously updates during drawing. Displayed measures include Stroke Density, Cross-Scale Coherence Index (CCI), Drift Amplitude, Tempo Variability, Motif Entropy, Stabilization Latency, and Commitment. These metrics provide real-time indicators of interaction dynamics rather than assessments of artwork content or symbolic meaning.

At the bottom of the metrics panel is a live Cognitive Trajectory visualization. This display represents the evolving interaction process as a trajectory through a multidimensional state space defined by exploration, coherence, drift, regulation, and stability. During active drawing, the visualization updates continuously as new interaction states are generated.

The interface also supports replay-based analysis, allowing researchers to revisit completed sessions, inspect organizational transitions, review detected events, and examine larger trajectory chapters. Together, these components transform drawing activity into a continuously evolving representation of creative engagement.

The figure illustrates the core design philosophy of the CTL: the completed artwork is not treated as the sole object of analysis. Instead, the system continuously captures, quantifies, and visualizes the unfolding dynamics of participation, enabling process-oriented investigation of creativity, regulation, exploration, and therapeutic change through time.
}
  \label{fig:cogLab}
\end{figure*}

\section{The Cognitive Trajectory Laboratory: Quantifying Creative Engagement Through Time}

To operationalize the trajectory-based framework proposed throughout this paper, we developed the \textit{Cognitive Trajectory Laboratory} (CTL), a research-oriented instrumented drawing environment designed to capture, analyze, and interpret creative engagement as a temporally unfolding process. Unlike traditional drawing applications, the CTL is not primarily intended as a creative production tool. Rather, it functions as an observational and analytical environment capable of transforming raw interaction traces into cognitive trajectories, trajectory properties, organizational events, and process-oriented interpretations.

\begin{figure*}[H]
  \centering
  \includegraphics[width=\textwidth]{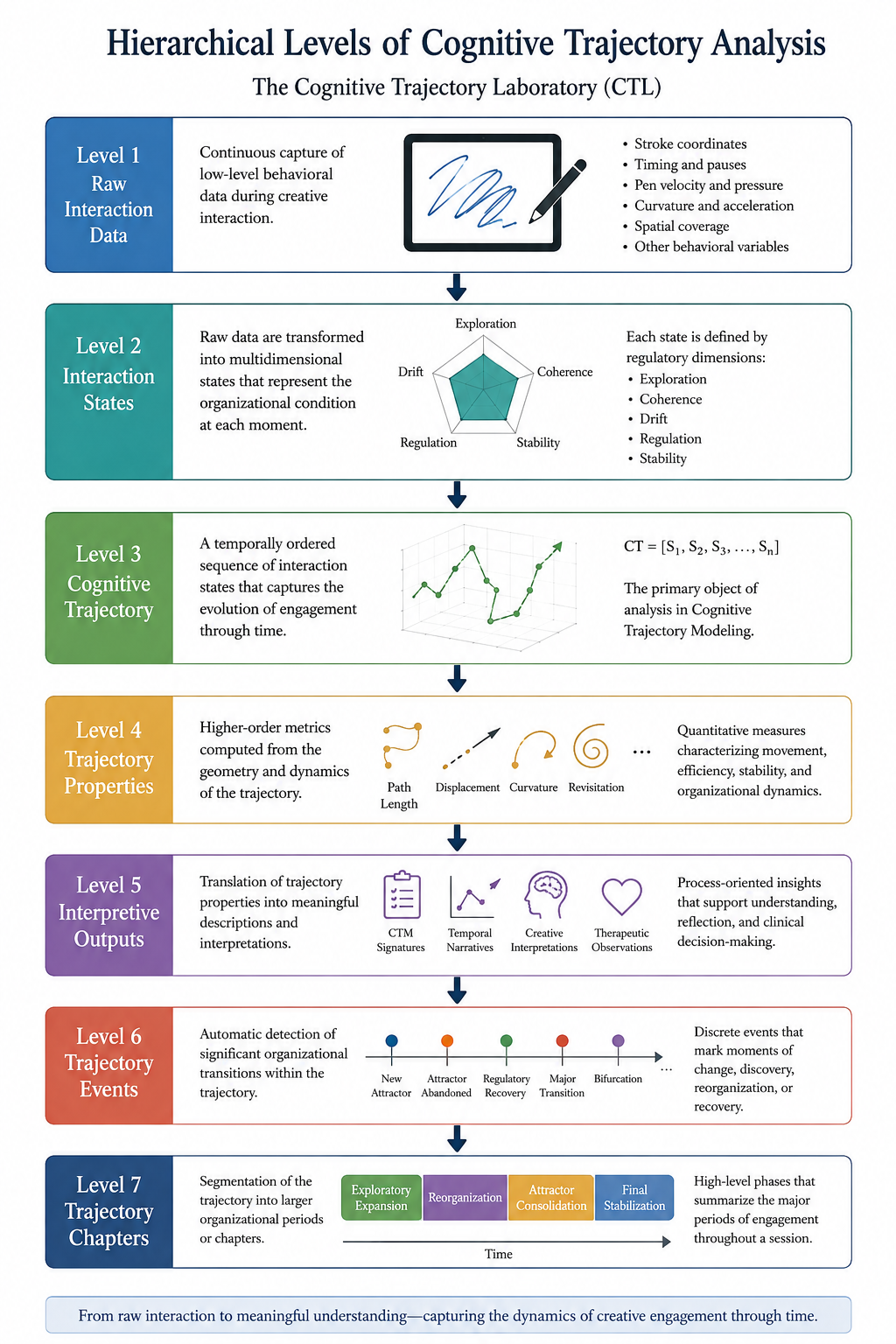}
  \caption{\textbf{Hierarchical Levels of Cognitive Trajectory Analysis in the Cognitive Trajectory Laboratory (CTL)}, illustrating the transformation of raw interaction traces into cognitive trajectories, trajectory properties, interpretive outputs, trajectory events, and higher-order trajectory chapters.}
  \Description{Description:
This figure presents the seven hierarchical levels of Cognitive Trajectory Analysis implemented within the Cognitive Trajectory Laboratory (CTL). The framework transforms continuous behavioral interaction data into progressively higher-order representations of creative engagement and therapeutic process.

Level 1: Raw Interaction Data consists of continuously recorded behavioral variables generated during drawing activity, including stroke coordinates, timing, pauses, velocity, curvature, acceleration, spatial coverage, and related interaction measures.

Level 2: Interaction States transforms raw behavioral data into multidimensional organizational states. Each state is defined by regulatory dimensions including exploration, coherence, drift, regulation, and stability, providing a momentary description of the interaction process.

Level 3: Cognitive Trajectory represents engagement as a temporally ordered sequence of interaction states. The trajectory serves as the primary unit of analysis in Cognitive Trajectory Modeling and captures how engagement evolves through time.

Level 4: Trajectory Properties derives quantitative characteristics from the geometry and dynamics of the trajectory. Examples include path length, displacement, curvature, revisitation, stability, efficiency, and other measures describing organizational movement through state space.

Level 5: Interpretive Outputs translates trajectory properties into process-oriented summaries and representations, including CTM signatures, temporal narratives, creative interpretations, and therapeutic observations.

Level 6: Trajectory Events identifies discrete moments of organizational significance within the trajectory, such as the discovery of new attractors, attractor abandonment, regulatory recovery, major transitions, and trajectory bifurcations.

Level 7: Trajectory Chapters segments the trajectory into larger periods of organization that summarize major phases of engagement. The example shown includes Exploratory Expansion, Reorganization, Attractor Consolidation, and Final Stabilization.

Together, the hierarchy illustrates how the CTL progresses from direct behavioral observations to increasingly abstract representations of creative process. Rather than focusing exclusively on completed artworks, the framework models the unfolding dynamics of engagement through time, providing a process-oriented methodology for studying creativity, regulation, and therapeutic change.}
  \label{fig:cogLab}
\end{figure*}

The laboratory extends earlier work on instrumented drawing systems by implementing a hierarchical architecture for Cognitive Trajectory Modeling (CTM) \cite{davis2026ctm}. Rather than treating interaction traces as isolated behavioral observations, the CTL organizes creative activity into multiple levels of representation, each corresponding to a different degree of abstraction. This hierarchical approach enables the transformation of low-level behavioral data into higher-level descriptions of cognitive organization, regulation, exploration, and adaptation.

Importantly, the system is intentionally non-interpretive. It does not attempt to infer emotional states, diagnose psychological conditions, or determine the symbolic meaning of artwork. Instead, it measures observable characteristics of interaction such as temporal organization, structural coherence, exploratory behavior, regulatory transitions, and trajectory dynamics. The laboratory therefore provides a process-oriented environment for studying creative regulation and therapeutic engagement without replacing clinical judgment or qualitative interpretation.

A live research prototype of the Cognitive Trajectory Laboratory is available at \url{https://www.enactivearttherapy.com}. The prototype is provided for demonstration and research purposes and should not be interpreted as a validated clinical assessment instrument.

\begin{figure*}[ht]
  \centering
  \includegraphics[width=\textwidth]{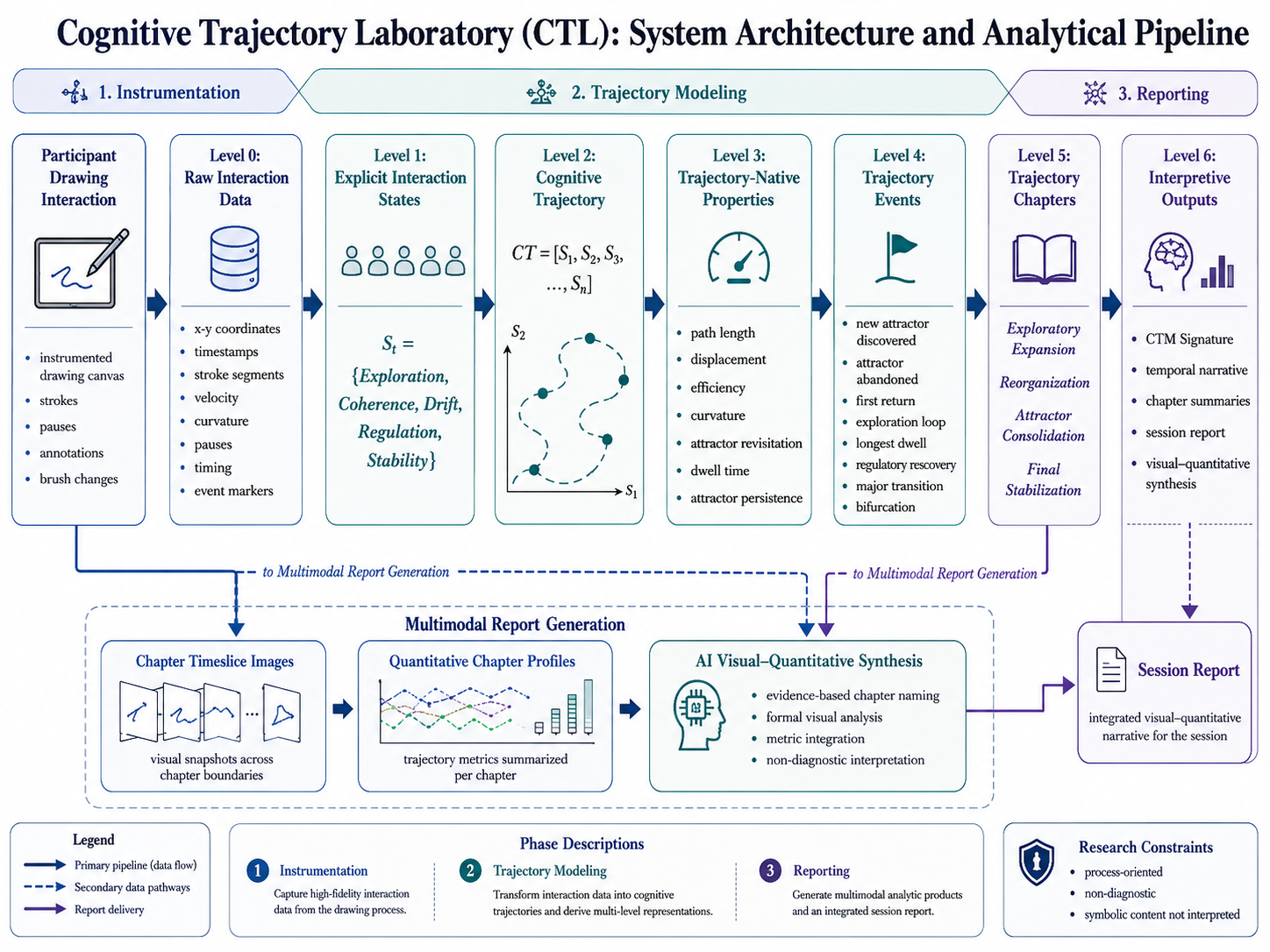}
  \caption{\textbf{System architecture and analytical pipeline of the Cognitive Trajectory Laboratory (CTL).}
  \normalfont{The framework transforms instrumented drawing interactions into cognitive trajectories, derives trajectory properties and events, segments trajectories into chapters, and generates integrated visual–quantitative reports through a multi-stage process of instrumentation, trajectory modeling, and reporting.}}
  \Description{Description:
This figure illustrates the complete system architecture and analytical pipeline of the Cognitive Trajectory Laboratory (CTL). The framework is organized into three major phases: Instrumentation, Trajectory Modeling, and Reporting.

The process begins with participant drawing interaction on an instrumented digital canvas. During drawing, the system continuously records interaction traces including strokes, pauses, annotations, brush changes, spatial coordinates, timing information, velocity, curvature, and other behavioral variables.

These interaction traces are stored as Level 0: Raw Interaction Data, which forms the foundation for subsequent analysis. Raw data are transformed into Level 1: Explicit Interaction States, where each moment of activity is represented by a multidimensional state defined by exploration, coherence, drift, regulation, and stability.

Sequences of interaction states are then organized into Level 2: Cognitive Trajectories, represented as temporally ordered paths through a multidimensional interaction space. Cognitive trajectories serve as the primary unit of analysis in Cognitive Trajectory Modeling (CTM).

From these trajectories, the system derives Level 3: Trajectory-Native Properties, including path length, displacement, efficiency, curvature, attractor revisitation, dwell time, and attractor persistence. These measures characterize the geometry and dynamics of engagement through time.

The system then identifies Level 4: Trajectory Events, which represent significant organizational transitions within the trajectory. Example events include new attractor discovery, attractor abandonment, first return to a prior attractor, exploration loops, regulatory recovery, major transitions, and trajectory bifurcations.

Detected events support the construction of Level 5: Trajectory Chapters, which segment the trajectory into larger organizational periods. The example shown includes Exploratory Expansion, Reorganization, Attractor Consolidation, and Final Stabilization. These chapters summarize major phases of engagement across the session.

The highest level of analysis consists of Level 6: Interpretive Outputs, including CTM signatures, temporal narratives, chapter summaries, session reports, and integrated visual–quantitative syntheses.

A secondary reporting pipeline combines chapter-level trajectory information with visual snapshots of the artwork. Chapter timeslice images, quantitative chapter profiles, and trajectory-derived metrics are integrated through a multimodal report-generation process. An optional AI-assisted visual–quantitative synthesis generates evidence-based chapter descriptions, visual analyses, metric integration, and non-diagnostic narrative summaries. These outputs are assembled into a final session report.

The figure illustrates the complete transformation from raw interaction traces to higher-order process representations and report generation. Rather than treating the completed artwork as the sole object of analysis, the CTL models the evolving dynamics of creative engagement across multiple levels of organization, enabling process-oriented investigation of creativity, regulation, and therapeutic change through time.}
  \label{fig:cogLab}
\end{figure*}

\subsection{Instrumented Drawing Environment}

Figure~\ref{fig:cogLab} presents the primary interface of the Cognitive Trajectory Laboratory. The central canvas provides a space for free-form drawing while a collection of real-time metrics summarize evolving interaction dynamics. The environment continuously records spatial, temporal, and organizational characteristics of drawing behavior. Rather than analyzing completed drawings, the system analyzes the process through which drawings emerge.

The current implementation is a single-user drawing environment and does not yet incorporate therapist participation or co-creative AI interaction. Its purpose is to demonstrate how interaction traces can be transformed into quantitative representations of creative engagement and cognitive organization. Figure \ref{fig:cogLab} shows the user interface of the Cognitive Trajectory Laboratory, featuring a drawing canvas on the left and the right-hand panel displays several real-time metrics intended to capture distinct dimensions of creative engagement. These live metrics are elaborated below.

\subsubsection{Stroke Density}

Stroke Density measures the rate of drawing activity over a rolling time window. Higher values indicate increased production activity, while lower values indicate pauses, reflection, or reduced engagement. From a trajectory perspective, changes in stroke density provide information regarding shifts in activity levels across time.

\subsubsection{Cross-Scale Coherence Index (CCI)}

The Cross-Scale Coherence Index measures the degree to which local, regional, and global structures remain organized relative to one another. High coherence values indicate that patterns observed at different scales remain mutually compatible and integrated, whereas lower values suggest fragmentation or loss of organizational consistency across scales.

\subsubsection{Drift Amplitude}

Drift Amplitude measures departures from previously established interaction patterns. High drift values indicate substantial changes in direction, structure, or organizational tendencies, whereas lower values indicate greater continuity with preceding activity. Within the regulatory framework developed in this paper, drift is interpreted as a measure of ongoing deviation from established trajectories rather than as an indicator of error.

\subsubsection{Tempo Variability}

Tempo Variability measures fluctuations in the timing of drawing actions. The metric is derived from inter-stroke intervals and captures variations in pacing across the drawing process. Stable tempo patterns indicate relatively consistent engagement rhythms, whereas elevated variability suggests changing modes of interaction.

\subsubsection{Motif Entropy}

Motif Entropy measures the diversity and distribution of recurring structural patterns within the drawing. Higher entropy values indicate greater diversity of forms and organizational structures, while lower values indicate increasing concentration around a smaller set of motifs.

\subsubsection{Stabilization Latency}

Stabilization Latency measures the time required for activity patterns to settle into relatively stable configurations following substantial changes or disruptions. Within trajectory analysis, stabilization events represent transitions from exploration toward consolidation.

\subsubsection{Commitment State}

The Commitment indicator summarizes the current balance between exploration and consolidation. Periods characterized by high variation and continued structural change are labeled as exploratory, whereas periods characterized by recurring motifs and increasing coherence indicate growing commitment to particular organizational structures.

\subsection{Therapeutic Process Indicators}
The Cognitive Trajectory Laboratory also includes a therapist-facing layer that translates lower-level CTM dynamics into a small set of clinically legible process indicators. These features are displayed prominently in the live interface and summarize five dimensions of creative engagement: Engagement, Flexibility, Persistence, Emotional Regulation, and Integration. Engagement estimates the degree of sustained participation in the session; Flexibility reflects the capacity to shift direction or reorganize activity; Persistence characterizes the ability to remain involved across difficulty or disruption; Emotional Regulation estimates recovery and stabilization following periods of increased drift or reduced coherence; and Integration reflects the extent to which local actions, emerging structures, and the session as a whole become organized into a coherent process. An overall Process score provides a composite summary of these dimensions while preserving the individual indicators for more detailed inspection. These indicators should therefore be regarded as operational hypotheses about process dynamics that require future empirical validation.

\begin{figure}[h]
  \centering
  \includegraphics[width=.35\textwidth]{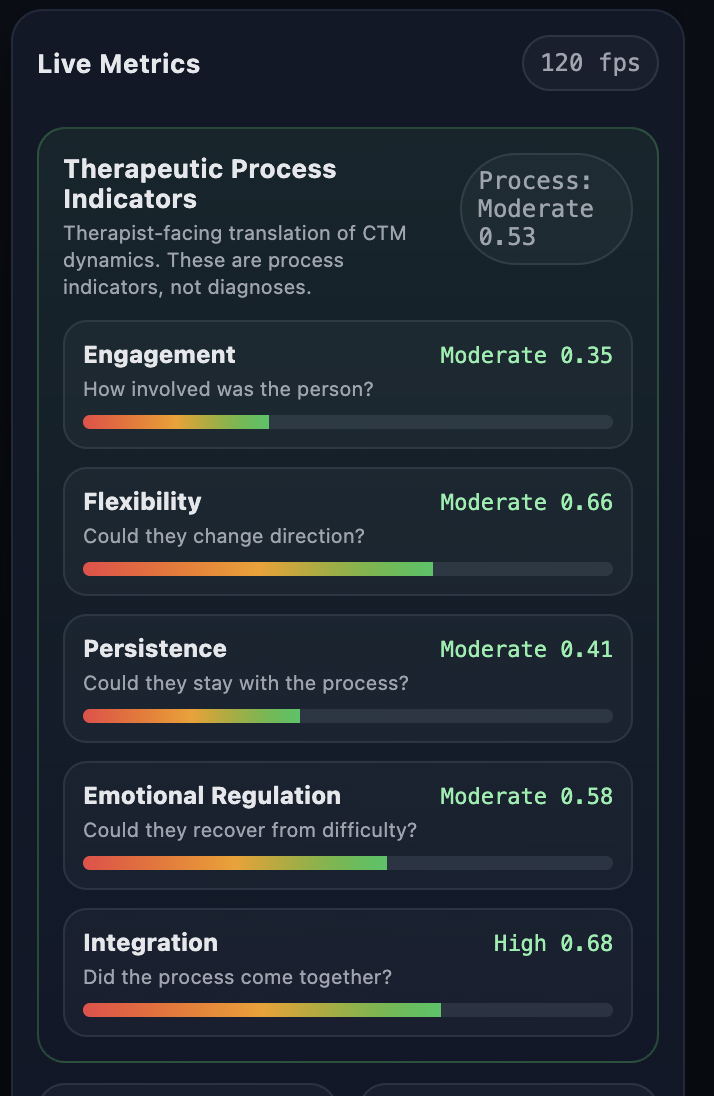}
  \caption{\textbf{Therapeutic Process Indicators in the Cognitive Trajectory Laboratory interface.}
\normalfont{The therapist-facing panel translates Cognitive Trajectory Modeling dynamics into five process indicators—Engagement, Flexibility, Persistence, Emotional Regulation, and Integration—along with an overall Process score. The indicators summarize observable patterns of participation and regulation to support process-oriented interpretation rather than diagnosis or inference of internal emotional states.}}
  \label{fig:heirarchy}
\end{figure}

These therapist-facing indicators are not direct measurements of internal emotional states, diagnostic categories, or therapeutic outcomes. Rather, they are quantitative translations of observable interaction dynamics derived from the participant’s temporal, spatial, structural, and trajectory-level behavior. Their purpose is to make the computational measures generated by the CTL more accessible to therapists and researchers by expressing them in terms that correspond to recognizable qualities of therapeutic participation. The indicators may therefore support reflection, session review, longitudinal comparison, and hypothesis generation, but they should be interpreted alongside clinical observation, participant self-report, contextual knowledge, and qualitative analysis rather than as independent clinical judgments.

\subsection{Trajectory Visualizations}

Whereas the metrics panel provides current values, the lower visualization panel reveals how those values evolve throughout the drawing session. Time-series visualizations display the trajectories of stroke density, coherence, drift, tempo variability, and motif entropy across time. Together these visualizations reveal changes in engagement intensity, structural organization, exploratory activity, and regulatory dynamics. The visualizations provide an intermediate representation between raw interaction traces and higher-level trajectory analyses. They reveal temporal patterns that may remain hidden within aggregate summary statistics while preserving the continuous evolution of engagement throughout a session.

\begin{figure*}[p]
  \centering
\includegraphics[width=.95\textwidth,height=.85\textheight,keepaspectratio]{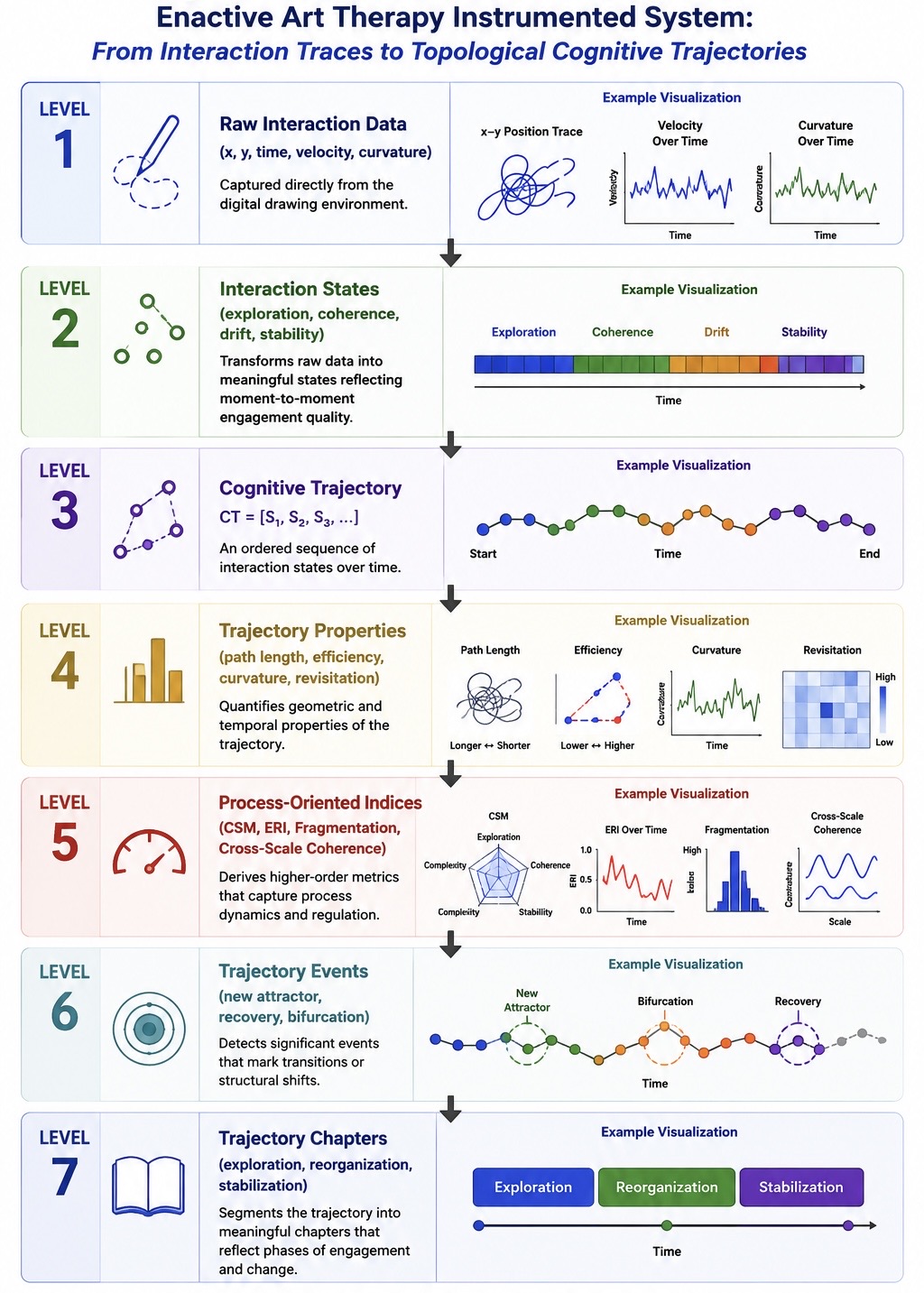}
  \caption{\textbf{Hierarchical levels of data within the Enactive Art Therapy instrumented system.} 
  \normalfont{The figure illustrates the progression from low-level behavioral interaction data to higher-order cognitive trajectories, process-oriented metrics, topological representations, and trajectory chapters. Each level contains representative variables and example visualizations generated by the Cognitive Trajectory Laboratory (CTL).}}
  \label{fig:heirarchy}
\end{figure*}

\subsection{Hierarchical Levels of Cognitive Trajectory Analysis}

The most significant contribution of the Cognitive Trajectory Laboratory is its hierarchical architecture for transforming interaction traces into cognitive trajectories and higher-order process representations. The CTL organizes creative activity across seven levels of analysis. This hierarchy, shown in Figure \ref{fig:heirarchy}, transforms CTM from a collection of interaction metrics into a multi-level framework for analyzing cognition, creativity, and regulation as temporally unfolding processes.

\subsection{Trajectory Properties and Events}

Once a Cognitive Trajectory has been constructed, a set of trajectory-native properties can be computed directly from its structure. Beyond global trajectory properties, the laboratory automatically identifies discrete moments of organizational significance. These events function as organizational markers that summarize important changes occurring within the evolving trajectory. Event detection enables researchers and clinicians to identify specific moments of stabilization, disruption, recovery, exploration, and reorganization rather than relying exclusively on aggregate measures.

\subsection{Replay and Retrospective Trajectory Analysis}

In addition to real-time monitoring, the Cognitive Trajectory Laboratory supports replay-based analysis. Complete interaction histories can be exported, imported, and reconstructed after a session has concluded. During replay, the original drawing process is synchronized with trajectory visualizations, trajectory chapters, and annotated events. Researchers and clinicians may therefore examine not only the final artwork but also the organizational dynamics that produced it. Annotation overlays, event filtering, chapter navigation, and retrospective trajectory inspection provide mechanisms for reviewing the temporal evolution of engagement in detail. This capability represents a significant methodological departure from traditional outcome-oriented assessment. Cognitive trajectories can be replayed, re-examined, and annotated after a session has concluded, enabling researchers and clinicians to inspect the temporal organization of engagement rather than relying solely on static outcomes.

Collectively, these levels transform the Cognitive Trajectory Laboratory from a system for computing interaction metrics into a hierarchical process-analysis architecture. Raw interaction traces become interaction states, interaction states become trajectories, trajectories generate properties and events, and events are organized into larger chapters and interpretive narratives. The laboratory therefore operationalizes Cognitive Trajectory Modeling as a multi-level methodology for investigating creativity, regulation, and therapeutic engagement as temporally unfolding processes.

\section{Cognitive Trajectory Reports}

\begin{figure*}[h]
  \centering
  \includegraphics[width=.85\textwidth]{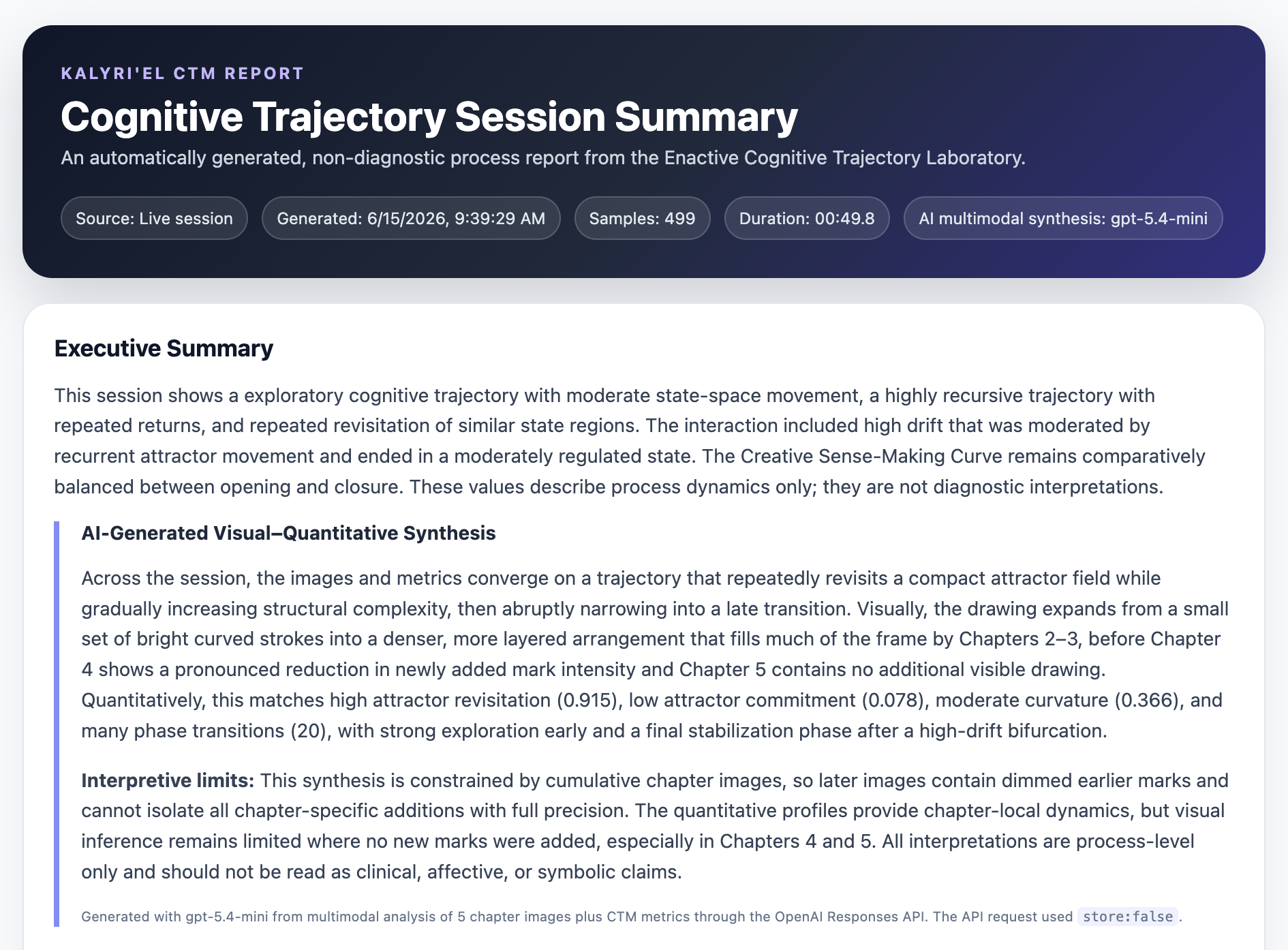}
  \caption{\textbf{CTL Report Introduction.}
\normalfont{The AI generates a narrative process-oriented description of the fine grained data as well as more abstract relational data generated by the CTL.}}
  \label{fig:summaryIntro}
\end{figure*}

The Cognitive Trajectory Laboratory generates a multi-level report (see Figure \ref{fig:summaryIntro}) that translates raw interaction traces into increasingly abstract representations of creative engagement. The resulting report is not intended to diagnose psychological conditions or infer symbolic meaning from artwork. Instead, it provides a process-oriented description of how participation unfolded through time. The report includes a deterministic executive summary generated from CTM variables and an optional AI-assisted multimodal synthesis derived from trajectory metrics and chapter-level images. The AI component does not generate the underlying measurements and is not treated as an independent clinical interpretation. Rather, it translates already-computed process features into a readable narrative while explicitly reporting interpretive limits. Because the synthesis has not yet been evaluated for agreement with expert raters, it should be regarded as an exploratory reporting aid rather than a validated analytic output.

\begin{figure*}[ht]
  \centering
  \includegraphics[width=\textwidth]{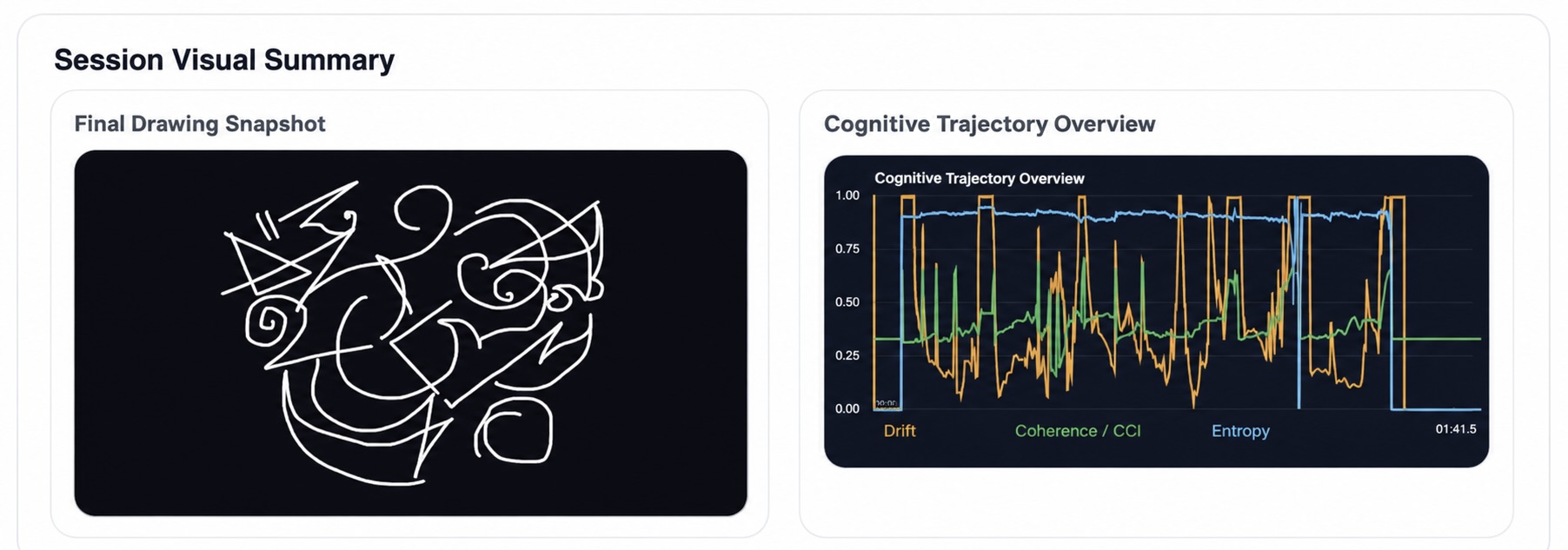}
  \caption{\textbf{Session visual summary generated by the Cognitive Trajectory Laboratory (CTL).} 
  \normalfont{The figure juxtaposes the completed drawing artifact with a temporal overview of the corresponding cognitive trajectory. The chart on the right plots three distinct values over time. Each colored line represents a different cognitive trajectory, as indicated by the labels beneath the chart: Drift (orange), Coherence / CCI (green), and Entropy (blue). This visualization illustrates how process-oriented measures complement traditional artifact-based representations of creative engagement.}}
  \Description{Description:

This figure presents a session-level visual summary generated by the Cognitive Trajectory Laboratory (CTL). The summary combines two complementary views of a drawing session: the completed artwork and a temporal representation of the interaction dynamics that occurred during its creation.

The left panel, labeled Final Drawing Snapshot, displays the completed drawing produced during the session. The artwork consists of a collection of interconnected curved and linear marks arranged across the drawing surface. This panel represents the final artifact traditionally available for assessment following completion of the creative activity.

The right panel, labeled Cognitive Trajectory Overview, presents a time-series visualization of process-oriented measures recorded during the same session. Multiple trajectory variables are displayed continuously across the duration of the interaction. The visualization illustrates fluctuations in organizational dynamics over time, including changes associated with exploration, coherence, drift, and related trajectory measures. Peaks, transitions, and periods of stability indicate shifts in the organizational structure of engagement as the session unfolds.

Together, the two panels illustrate a central distinction within the proposed framework. Traditional assessment approaches often focus primarily on the completed artwork as the principal object of analysis. In contrast, Cognitive Trajectory Modeling treats the creative process itself as an analyzable phenomenon. The completed drawing provides information about the final outcome of engagement, whereas the cognitive trajectory overview provides information about how that outcome emerged through time.

The figure demonstrates the complementary relationship between artifact-based and process-based representations of creative activity. Rather than replacing the artwork, the trajectory visualization provides an additional layer of analysis that captures temporal dynamics, organizational transitions, and patterns of engagement that are not directly observable from the final drawing alone.}
  \label{fig:visualization}
\end{figure*}

\subsection{Regulatory Dynamics}

The report summarizes aggregate characteristics of the session, including exploration, coherence, regulation, drift, and stability. Together, these measures provide a process-oriented description of how engagement evolved through time. Rather than evaluating the completed artwork alone, Regulatory Dynamics characterize the organizational tendencies that emerged during participation and how those tendencies changed throughout the creative process.

Exploration reflects the degree to which activity expanded into novel regions of interaction, while coherence estimates the extent to which emerging structures remained organized across time. Drift captures departures from previously established patterns and provides a measure of ongoing reorganization and change. Regulation describes the capacity of the interaction to maintain or restore viable engagement in the presence of uncertainty, disruption, or increasing complexity. Stability reflects the persistence of organizational structure across the session and provides an indication of whether engagement remained coherent despite local fluctuations and exploratory departures.

Taken together, these measures provide a high-level summary of the participant's evolving interaction with the creative medium. Rather than treating creativity as a static outcome, Regulatory Dynamics characterize the balance between exploration and consolidation, disruption and recovery, and change and continuity that unfolds throughout the art-making process.

\subsection{Trajectory Events}

While trajectory properties describe the session globally, trajectory events identify specific moments of organizational significance. These events function as markers of organizational change. For example, a major state-space transition indicates movement between distinct organizational regions, whereas a regulatory recovery indicates successful restoration of coherence following elevated drift. Importantly, trajectory events are not interpretations imposed upon the drawing. Rather, they are computed directly from changes in trajectory organization.

\subsection{Trajectory Chapters}

\begin{figure*}[h]
  \centering
  \includegraphics[width=.8\textwidth]{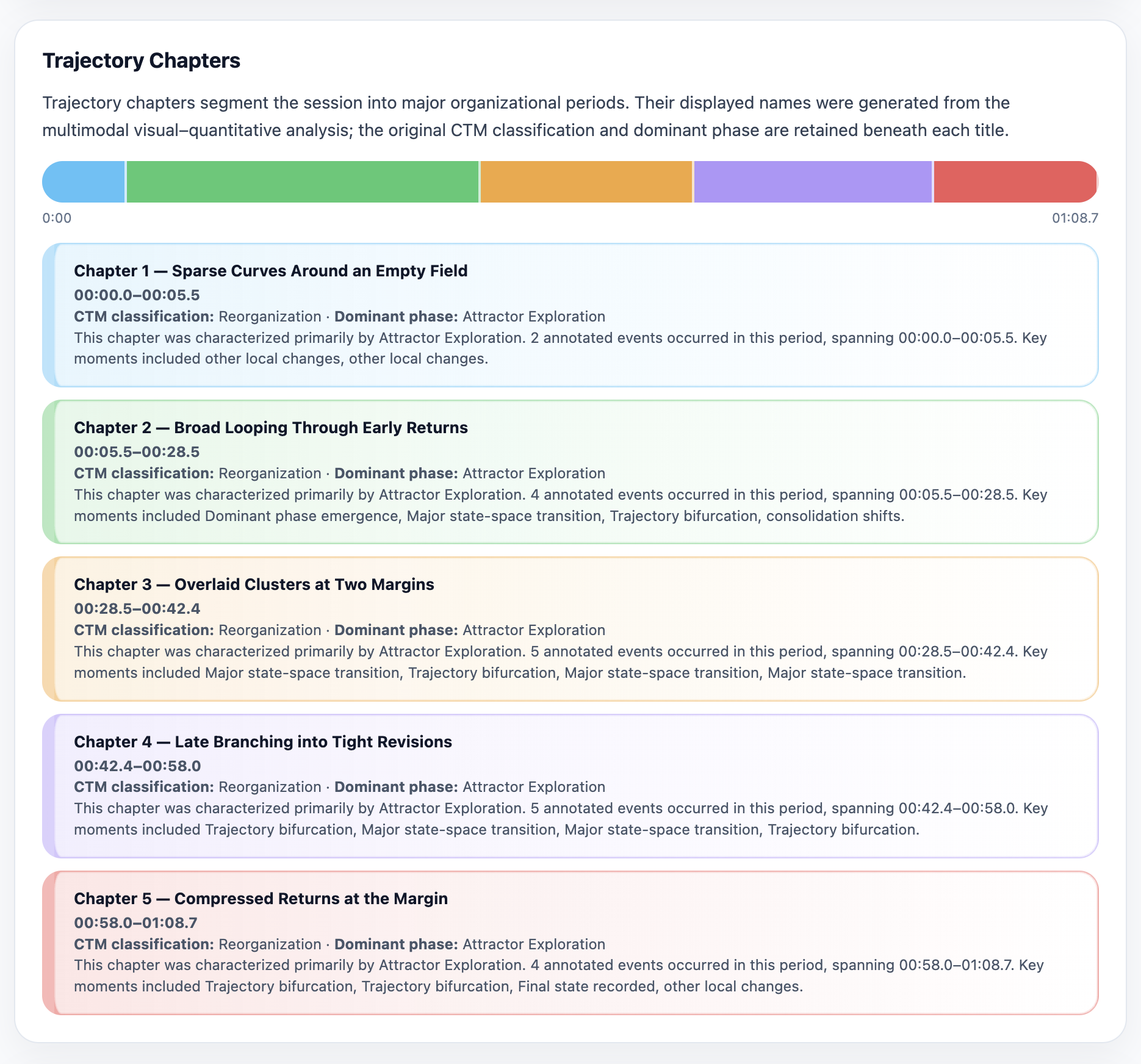}
  \caption{\textbf{Example trajectory chapter analysis generated by the Cognitive Trajectory Laboratory (CTL).} 
  \normalfont{The session is segmented into five major organizational periods, each summarized by a chapter title, temporal boundaries, dominant regulatory phase, CTM classification, and significant trajectory events. Together, the chapters provide a higher-level account of how engagement evolves throughout the session.}}
  \label{fig:chapters}
\end{figure*}

At the highest organizational level, the system segments the trajectory into chapters representing extended periods of relatively coherent activity, as shown in Figure \ref{fig:chapters}. For the example session, five major chapters were identified. Trajectory chapters provide a narrative description of organizational development without relying on symbolic interpretation. Rather than asking what the drawing represents, chapter analysis asks how the interaction evolved, as shown in Figure \ref{fig:chapters}. This distinction is central to the Cognitive Trajectory approach. The report does not infer emotional meaning, personality characteristics, or therapeutic outcomes from the artwork itself. Instead, it describes the organizational dynamics through which participation unfolded.

\begin{figure*}[h]
  \centering
  \includegraphics[width=\textwidth]{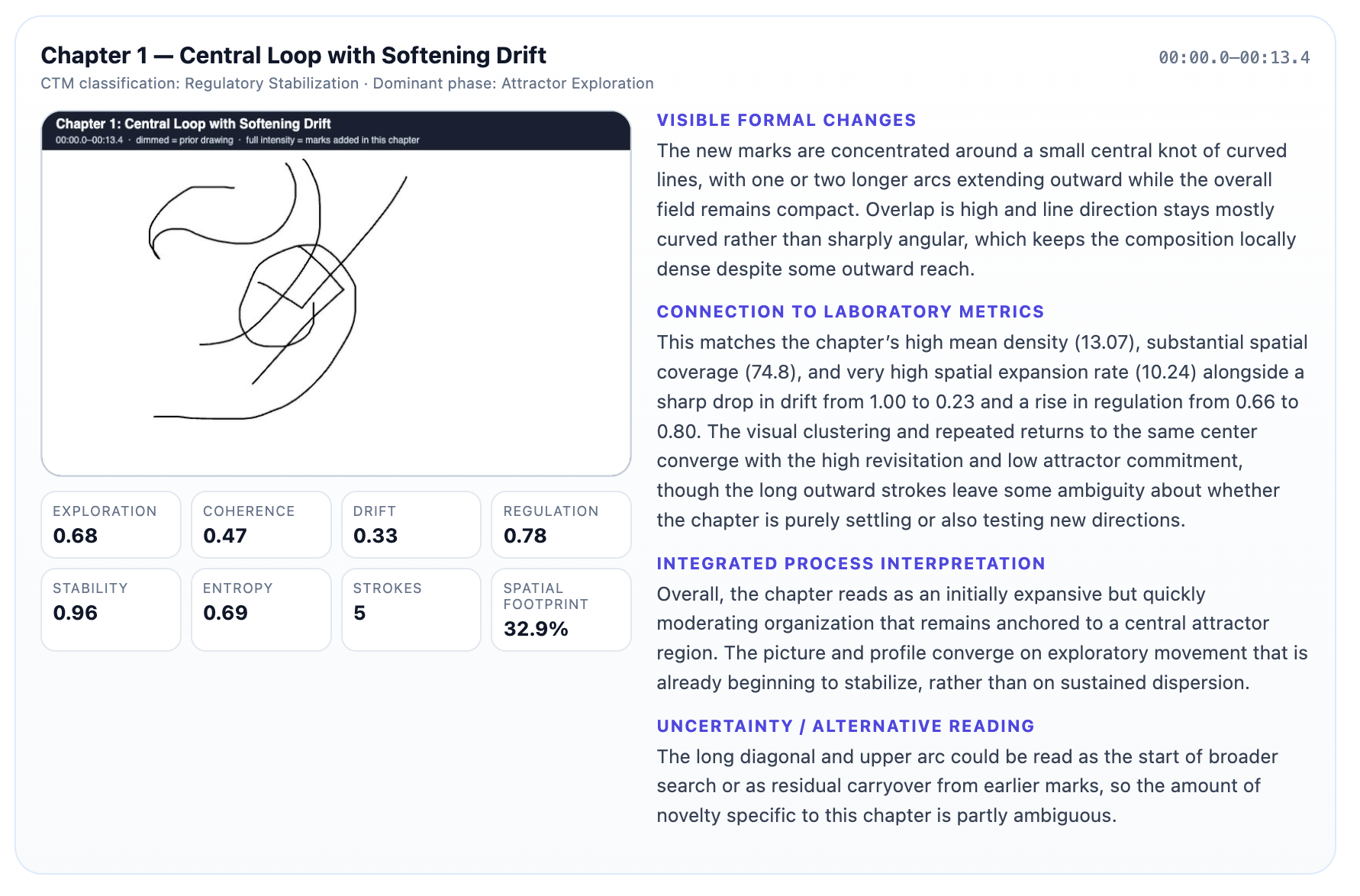}
  \caption{\textbf{Chapter-level analysis within a Cognitive Trajectory Report.}
\normalfont{The reporting framework segments a drawing session into temporally distinct trajectory chapters and generates integrated visual, quantitative, and interpretive summaries for each segment. By combining interaction metrics, trajectory dynamics, and image-based analysis, the report supports retrospective examination of how creative engagement unfolds across successive phases of activity.}}

  \label{fig:chapterTimeSlice}
\end{figure*}

\subsection{A Process-Oriented Interpretation}

From a Cognitive Trajectory perspective, the significance of creative engagement lies not primarily in the final image produced but in the dynamic process through which participation unfolds across time. Cognitive trajectories provide a process-oriented representation of this activity by revealing patterns of exploration, regulation, reorganization, stabilization, recovery, and change as they emerge during interaction. Rather than reducing creative engagement to isolated actions or completed outcomes, trajectory analysis examines how organizational structure develops, transforms, and persists throughout the creative process.

The resulting report serves as a quantitative description of creative engagement itself rather than an interpretation of symbolic content or psychological meaning. By transforming interaction traces into trajectories, trajectory properties, events, chapters, and process-oriented indices, the framework provides multiple levels of representation through which the evolving dynamics of participation can be examined. This approach shifts assessment from static products and retrospective accounts toward the observable organization of engagement as it unfolds through time.

The Cognitive Trajectory Laboratory has been implemented as an interactive research prototype. The current system, supporting instrumented drawing, trajectory visualization, replay, data export, and report generation, is available at \url{https://www.enactivearttherapy.com/}.

\section{Process-Oriented Dimensions of Creative Engagement}

While trajectory properties characterize the geometry and organization of cognitive trajectories, the Cognitive Trajectory Laboratory also computes a collection of higher-level process indices designed to summarize the evolving dynamics of creative engagement. These measures occupy an intermediate level between raw trajectory analysis and interpretive outputs. Rather than describing isolated behavioral variables, they quantify recurring patterns of exploration, regulation, coherence, commitment, and organizational change that emerge across extended periods of interaction. Together, these indices provide a process-oriented account of creative engagement. Rather than evaluating what a drawing represents, they describe how participation unfolds through time. The resulting measures make it possible to examine regulatory flexibility, exploratory behavior, stabilization, and organizational transformation as observable properties of creative activity.

\subsection{Creative Sense-Making Curve}

\begin{figure*}[h]
  \centering
  \includegraphics[width=.8\textwidth]{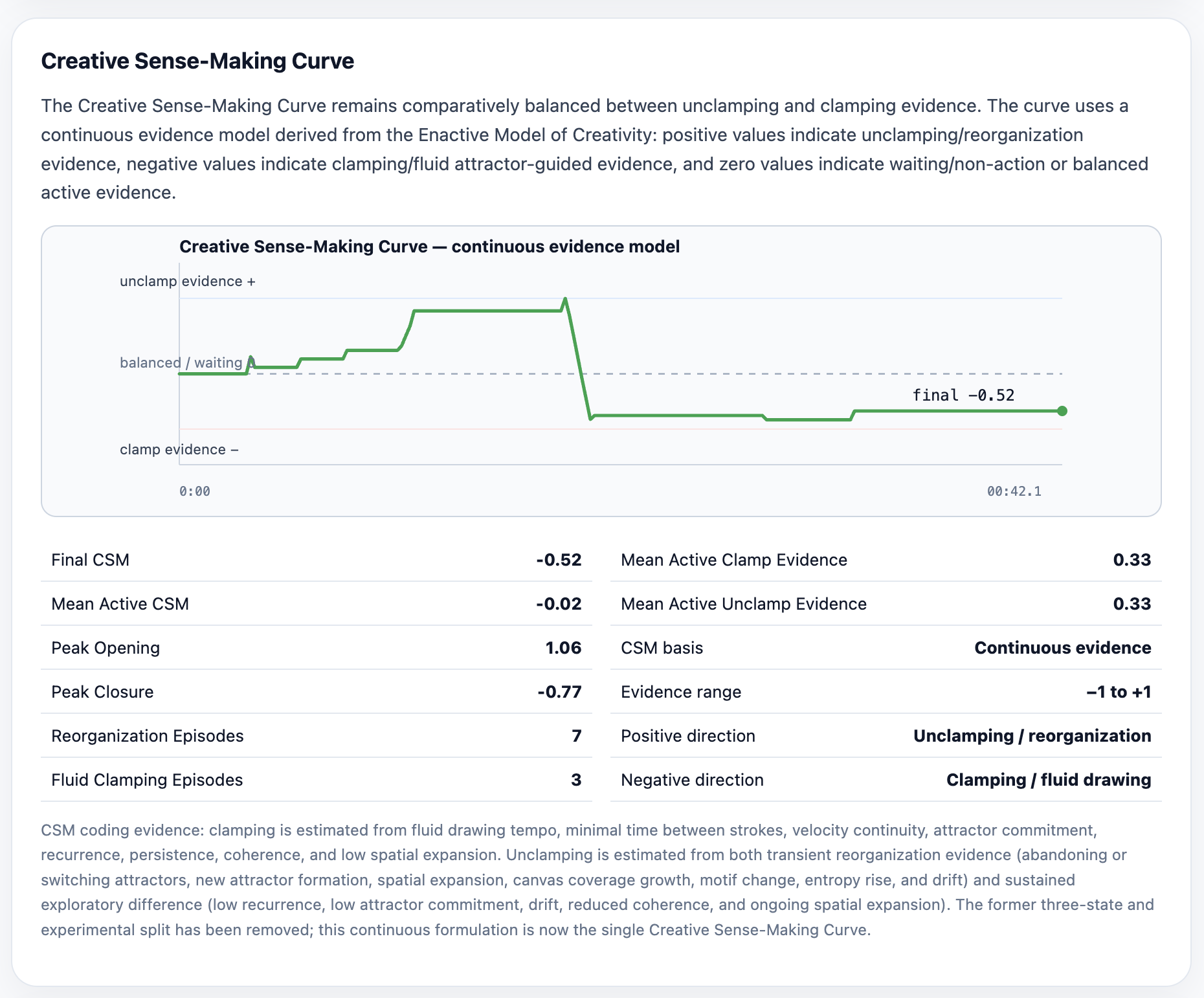}
  \caption{\textbf{Creative Sense-Making (CSM) Curve generated from a cognitive trajectory.}
  \normalfont{The curve represents a continuous evidence model derived from the Enactive Model of Creativity, where positive values indicate evidence for unclamping and reorganization, negative values indicate evidence for clamping and fluid attractor-guided engagement, and values near zero indicate balanced or transitional conditions.}}
  \Description{Description:

This figure presents a Creative Sense-Making (CSM) Curve, a process-oriented representation derived from Cognitive Trajectory Modeling and the Enactive Model of Creativity. The curve summarizes the balance between organizational opening and organizational closing throughout a creative session.

The vertical axis represents a continuous evidence scale ranging from negative to positive values. Positive values correspond to unclamping evidence, indicating reorganization, increased openness, exploratory restructuring, and movement away from established attractors. Negative values correspond to clamping evidence, indicating consolidation, attractor-guided activity, stabilization, and the preservation of existing organizational structures. Values near zero indicate relatively balanced conditions in which neither opening nor closing tendencies dominate.

The horizontal axis represents elapsed session time. The green line traces the evolving balance between opening and closing evidence throughout the interaction. Rather than categorizing behavior into discrete states, the curve provides a continuous account of how organizational tendencies fluctuate across the session.

In the example shown, the trajectory remains close to the balanced region for most of the interaction. Early portions of the session exhibit mild clamping evidence, followed by movement toward a more balanced state and brief periods of weak unclamping evidence. Later portions of the session return toward balanced conditions before ending with a modest negative value. The final CSM value of −0.18 indicates a slight predominance of clamping evidence at the conclusion of the session.

The summary statistics beneath the curve provide aggregate measures of the session. These include the final CSM value, mean active CSM, peak opening, peak closure, the number of reorganization episodes, the number of fluid clamping episodes, mean active clamp evidence, and mean active unclamp evidence. Together, these measures characterize the overall balance between exploratory reorganization and organizational consolidation across the interaction.

The Creative Sense-Making Curve is intended to describe process dynamics rather than psychological states, emotions, or symbolic content. Within the Cognitive Trajectory Laboratory, the curve functions as a higher-order summary of regulatory organization, providing a continuous representation of how engagement alternates between opening and closing tendencies throughout creative activity. The visualization supports the broader goal of modeling creativity as an evolving process of organizational regulation rather than a static property of the completed artifact.}
  \label{fig:csmCurve}
\end{figure*}

The Creative Sense-Making Curve provides a continuous representation of the balance between exploratory opening and organizational closure throughout a drawing session, as shown in Figure \ref{fig:csmCurve}. Derived from the Enactive Model of Creativity and Creative Sense-Making theory, the measure estimates the relative strength of evidence for reorganization versus consolidation at each moment in time. Positive values indicate evidence for exploratory opening, including spatial expansion, attractor transitions, motif diversification, organizational drift, and the emergence of novel structures. Negative values indicate evidence for organizational closure, including recurrence, attractor commitment, repeated motif usage, coherent continuation, and fluid drawing within established organizational regions. Unlike traditional outcome measures, the CSM curve preserves the temporal structure of creative engagement. The resulting trajectory reveals how participants oscillate between exploration and consolidation as creative activity unfolds. Aggregate measures such as Final CSM, Mean Active CSM, Peak Opening, Peak Closure, and counts of reorganization and clamping episodes summarize these dynamics at the session level.

\subsection{Exploration–Regulation Index (ERI)}

The Exploration–Regulation Index (ERI) quantifies the relationship between exploratory behavior and regulatory organization. Creative engagement often requires movement beyond established structures while simultaneously maintaining sufficient coherence to remain viable. ERI was developed to capture this balance. Higher ERI values indicate exploratory engagement that remains effectively regulated and organized. Lower values indicate either excessive rigidity, in which exploration is suppressed, or excessive drift, in which exploratory activity overwhelms organizational stability. ERI therefore serves as a measure of adaptive creative flexibility and provides a quantitative estimate of how successfully exploration and regulation are coordinated throughout a session.

\subsection{Fragmentation Index}

The Fragmentation Index estimates the degree to which interaction becomes dispersed across multiple competing organizational tendencies. High fragmentation indicates increasing separation among interaction patterns, attractors, or regulatory states, suggesting reduced integration across the evolving trajectory. Fragmentation is not necessarily maladaptive. Temporary increases often accompany periods of exploration and reorganization. However, sustained fragmentation may indicate difficulty maintaining coherent participation. The index therefore provides a useful complement to measures of coherence and stability.

\subsection{Cross-Scale Coherence Dynamics}

Creative engagement unfolds simultaneously across multiple scales of organization. Local drawing actions combine to form regional structures, which in turn contribute to larger global patterns. Cross-Scale Coherence Dynamics measure the degree to which these levels remain mutually compatible throughout a session. High coherence indicates that local, regional, and global organizational structures remain integrated despite ongoing change. Lower coherence indicates increasing divergence among scales and may signal emerging fragmentation or reorganization. By examining coherence dynamically rather than as a static property, the CTL can identify periods of organizational integration, disruption, and recovery.

\subsection{Attractor Commitment and Landscape Formation}

As interaction unfolds, participants often revisit particular regions of state space and gradually commit to recurring organizational patterns. Attractor Commitment measures the extent to which engagement becomes concentrated around established organizational structures. Closely related is Landscape Formation, which estimates the degree to which stable organizational regions have emerged within the evolving state space. Together, these measures provide insight into how exploratory activity gives rise to recurring structures and how participants balance novelty with commitment throughout the creative process.

\section{Discussion and Implications for Art Therapy Research}
The framework developed in this paper positions art therapy as a measurable process of cognitive and interactional change. Rather than focusing exclusively on completed artworks, symbolic interpretation, or therapeutic outcomes, the framework directs attention toward the evolving dynamics of creative engagement itself. This shift has important implications for research methodology, theory development, and the future integration of computational approaches within art therapy.

Contemporary art therapy research has increasingly emphasized the importance of identifying mechanisms of change rather than solely documenting therapeutic outcomes. Although substantial evidence supports the effectiveness of art therapy and other creative arts therapies across a variety of clinical contexts, researchers continue to call for process-oriented investigations capable of explaining how therapeutic change emerges during creative engagement.
The trajectory-based framework proposed here contributes to this effort by providing a set of concepts and measures through which creative processes can be observed, represented, and analyzed across time.
\subsection{From Outcome Measurement to Process Measurement}
Traditional assessment approaches often evaluate change through comparisons between pre-treatment and post-treatment states. Such methods can reveal whether improvement has occurred but provide relatively little information about the processes through which change emerges.
A trajectory-based perspective shifts attention from static outcomes toward ongoing dynamics. Creative activity is treated as a continuously evolving process characterized by stabilization, disruption, exploration, recovery, and reorganization. By examining these processes directly, researchers gain access to aspects of therapeutic change that remain largely invisible within outcome-based approaches.
This perspective aligns with broader calls within creative arts therapies for increased investigation of change processes and mechanisms of action. Rather than asking only whether art therapy works, researchers can begin asking how therapeutic change unfolds through creative engagement itself.

\subsection{Longitudinal Analysis of Therapeutic Change}
The temporal nature of cognitive trajectories also supports longitudinal research. Therapeutic change rarely occurs as a linear progression. Individuals often experience periods of growth, disruption, consolidation, and reorganization across multiple sessions.
Because cognitive trajectories preserve the temporal structure of engagement, they provide a means of examining how regulatory and creative capacities evolve across extended periods of time. Researchers can investigate whether changes in drift tolerance, coherence maintenance, exploratory behavior, or participatory engagement emerge gradually or through periods of rapid reorganization.
Such analyses may help bridge the gap between micro-level interaction processes and macro-level therapeutic outcomes. Rather than treating sessions as isolated events, researchers can examine how patterns accumulate and transform across the course of treatment.
\subsection{Computational Analysis of Creative Engagement}
The proposed framework also creates opportunities for computational approaches to art therapy research. Advances in digital art systems, interaction logging, and computational creativity make it increasingly possible to capture detailed traces of creative activity as it unfolds.
These interaction traces can be analyzed using methods drawn from dynamical systems theory, human-computer interaction, computational creativity, and cognitive modeling. Such approaches make it possible to quantify dimensions of engagement that are difficult to observe through traditional methods alone, including temporal organization, interactional influence, recovery dynamics, and shifts in participation patterns.
Importantly, the goal of computational analysis is not to replace qualitative interpretation or therapeutic expertise. Rather, computational methods provide additional tools for examining processes that may otherwise remain difficult to observe directly. In this sense, computational analysis serves as a complement to existing art therapy methodologies rather than an alternative to them.
\subsection{Future Research Questions}
The framework presented in this paper generates a number of empirically testable research questions.
One important question concerns prediction. If cognitive trajectories capture meaningful aspects of creative engagement, then particular trajectory patterns may be associated with therapeutic improvement. Future research may investigate whether changes in coherence, exploration, recovery, or participation predict clinical outcomes across different populations and intervention types.
A second question concerns the relationship between creativity and regulation. Contemporary art therapy theories frequently assume that creative activity contributes to emotional regulation and psychological well-being. However, the mechanisms linking creativity and regulation remain incompletely understood. Trajectory-based analyses may help clarify how exploratory behavior, coherence maintenance, and adaptive reorganization contribute to therapeutic change.
A third question concerns interaction. If creativity emerges through participation rather than solely within individuals, then successful therapy may depend upon particular patterns of interaction among clients, therapists, materials, and environments. Future studies may investigate which forms of reciprocity, participation balance, and creative contribution are associated with productive therapeutic engagement.
Taken together, these questions point toward a broader research program centered on the dynamics of creative interaction. Rather than treating creativity as a static outcome or individual trait, the present framework encourages researchers to investigate how therapeutic change emerges through evolving trajectories of participation, regulation, and sense-making.

\subsection{Validation Agenda}

Establishing the Cognitive Trajectory Laboratory as a reliable research instrument will require a staged program of technical, construct, convergent, and predictive validation. First, \textbf{technical reliability} should determine whether the system produces stable and reproducible outputs from identical or systematically controlled interaction traces. This includes testing metric determinism, sensitivity to sampling rate and device characteristics, robustness to missing or noisy input, and test–retest consistency under comparable drawing conditions. Such evaluation is especially important because CTL measures are generated through a multilevel computational pipeline in which variation at the level of raw interaction data may propagate into interaction states, trajectory properties, events, and chapters. Consistent with established measurement frameworks, reliability should therefore be evaluated separately from validity rather than treated as evidence that the constructs themselves are meaningful.

Second, \textbf{construct validity} should examine whether CTL variables behave in ways that are consistent with their theoretical definitions. Measures of exploration, coherence, drift, regulation, stability, attractor commitment, and fragmentation should be tested under experimental conditions designed to selectively elicit related interaction patterns. For example, open-ended prompts may be compared with highly constrained tasks, planned drawings with improvisational drawings, and uninterrupted sessions with sessions containing controlled perturbations. Construct validation should also establish that higher-order variables are not reducible to simpler behavioral quantities such as stroke count, session duration, drawing speed, or spatial coverage. Evidence would be strengthened if theoretically related measures covary while empirically distinct dimensions remain discriminable. Because validity concerns the interpretation and intended use of scores rather than the existence of a metric alone, validation claims should remain specific to particular populations, tasks, and research contexts.

Third, \textbf{convergent validity} should investigate whether trajectory-based measures correspond meaningfully with independent observations of the same or related processes. CTL outputs could be compared with blinded therapist ratings, participant self-reports, behavioral coding of recorded sessions, established measures of affect regulation and creative engagement, and physiological indicators such as heart-rate variability or electrodermal activity. Convergence should not require perfect agreement, because these methods operate at different descriptive levels. Rather, the goal would be to determine whether theoretically related measures exhibit interpretable associations while CTL continues to contribute process information not captured by static assessments or retrospective reports. Divergent or discriminant analyses should likewise test whether CTL variables remain distinguishable from unrelated constructs and from general motor activity.

Finally, \textbf{predictive and longitudinal validity} should test whether within-session and across-session trajectory patterns anticipate outcomes that matter for research or therapeutic practice. Candidate questions include whether regulatory recovery predicts sustained engagement, whether increasing exploration–regulation balance precedes improvement on independent measures, whether recurrent fragmentation predicts disengagement, and whether changes in trajectory organization across sessions correspond with therapist-rated or participant-reported change. Such studies should use preregistered hypotheses, held-out samples, and prospective designs to avoid interpreting descriptive patterns retrospectively as clinically meaningful. Where CTL-derived variables are used as digital endpoints, verification and validation should be conducted in relation to the specific context of use, with explicit attention to device performance, usability, data integrity, and the derivation of higher-order measures from raw sensor data.

Together, these stages would support a gradual transition from computational prototype to validated process-measurement environment. Validation should not seek to establish that a single CTL score directly reveals a psychological state. Instead, it should determine which trajectory measures are technically dependable, theoretically interpretable, empirically associated with independent evidence, and useful for explaining or predicting change within clearly defined contexts. Until such evidence is available, CTL outputs should be treated as research-oriented descriptions of interaction dynamics rather than diagnostic or clinical indicators.

\subsection{Toward a Trajectory Science of Art Therapy}
The broader implication of this work is the possibility of a process-oriented science of art therapy. Such an approach would complement existing symbolic, narrative, and neurobiological frameworks by providing methods for representing and analyzing the dynamics of engagement itself.
Cognitive trajectories, creative dynamics, drift, coherence, tempo regulation, and exploration measures offer a common vocabulary through which therapeutic processes can be studied across modalities, populations, and contexts. By making creative engagement measurable, the framework opens new opportunities for theory development, empirical investigation, and interdisciplinary collaboration among art therapy, creativity research, cognitive science, and computational methods.

\section{Conclusion}

Art therapy has traditionally been understood through the interpretation of images, narratives, and therapeutic outcomes. While these approaches remain valuable, they provide limited visibility into the dynamic processes through which creative engagement unfolds. This paper has argued that art therapy may also be understood as a measurable process of cognitive regulation occurring through ongoing interaction with materials, environments, and creative activity. To address this challenge, the paper introduced a trajectory-based framework for examining therapeutic change through time. Drawing upon enactive cognition, participatory sense-making, and dynamical systems theory, cognitive trajectories were proposed as temporally ordered representations of evolving engagement during creative activity. Concepts such as drift, coherence, tempo regulation, and exploration provide a foundation for quantifying aspects of creative participation that have traditionally remained difficult to observe directly. The instrumented drawing environment presented in this paper demonstrates how interaction traces can be transformed into process-oriented measures capable of supporting trajectory-based analysis. Ultimately, the central contribution of this work is a shift in perspective. Rather than treating art therapy solely as a medium for producing meaningful artifacts, the framework proposed here treats creative engagement itself as an observable and measurable process. By focusing on interaction traces and cognitive trajectories, future research may gain new tools for studying how therapeutic change emerges through the unfolding dynamics of creative participation.

\section*{Acknowledgments}

The author acknowledges the use of OpenAI's ChatGPT as a research and writing assistant during the development of this manuscript. Through an extended iterative dialogue, the system contributed to brainstorming, conceptual clarification, organizational refinement, editorial feedback, and the exploration of alternative formulations of theoretical arguments. The author remained responsible for all ideas, interpretations, theoretical claims, analyses, and final editorial decisions. The resulting manuscript emerged through a process of ongoing human--AI interaction in which AI-assisted reflection, critique, and revision supported the development and communication of the work.

\bibliographystyle{ACM-Reference-Format}
\bibliography{pubs2}


\end{document}